\documentclass[letterpaper]{article} 
\usepackage{aaai24}  
\usepackage{times}  
\usepackage{helvet}  
\usepackage{courier}  
\usepackage[hyphens]{url}  
\usepackage{graphicx} 
\usepackage{natbib}  
\usepackage{caption} 
\usepackage{subfigure}
\usepackage{epsfig}
\usepackage{makecell}
\usepackage{enumitem}
\usepackage{booktabs} 
\usepackage{amsfonts}   
\usepackage{amsmath}
\usepackage{nicefrac}       
\usepackage{microtype}      
\usepackage{xcolor}         
\usepackage{cases}
\usepackage{url}  
\usepackage{comment}
\usepackage{multirow}

\usepackage[ruled,linesnumbered]{algorithm2e}

\newtheorem{definition}{Definition}
\newtheorem{theorem}{Theorem}

\newcommand{\A}{\mathcal{A}}
\newcommand{\D}{\mathcal{D}}
\newcommand{\leftdollar}{\leftarrow}
\newcommand{\pke}{\textsf{\upshape{PKE}}}
\newcommand{\gen}{\textsf{\upshape{Gen}}}
\newcommand{\enc}{\textsf{\upshape{Enc}}}
\newcommand{\dec}{\textsf{\upshape{Dec}}}
\newcommand{\fake}{\textsf{\upshape{Fake}}}

\newcommand{\csl}{\mathsf{CS}_{\textup{lite}}}

\usepackage{newfloat}
\usepackage{listings}
\DeclareCaptionStyle{ruled}{labelfont=normalfont,labelsep=colon,strut=off} 
\title{EncryIP: A Practical Encryption-Based Framework for Model Intellectual Property Protection}
\author{
Xin Mu\textsuperscript{\rm 1}, 
Yu Wang\textsuperscript{\rm 1}, 
Zhengan Huang\textsuperscript{\rm 1}, 
Junzuo Lai\textsuperscript{\rm 2}, 
Yehong Zhang\textsuperscript{\rm 1}, 
Hui Wang\textsuperscript{\rm 1}, 
Yue Yu\textsuperscript{\rm 1}\thanks{Corresponding author.}
}
\affiliations{
\textsuperscript{\rm 1}Peng Cheng Laboratory, Shenzhen 518000, China\\
\textsuperscript{\rm 2}College of Information Science and Technology, Jinan University, Guangzhou, China\\

\{mux, wangy12, wangh06, yuy\}@pcl.ac.cn, \{zhahuang.sjtu, laijunzuo, zyhredleaf\}@gmail.com
}

\begin{document}

\maketitle

\begin{abstract}
	In the rapidly growing digital economy, protecting intellectual property (IP) associated with digital products has become increasingly important.  Within this context, machine learning (ML) models, being highly valuable digital assets, have gained significant attention for IP protection. 
	This paper introduces a practical encryption-based framework called \textit{EncryIP}, which seamlessly integrates a public-key encryption scheme into the model learning process. 
	This approach enables the protected model to generate randomized and confused labels, ensuring that only individuals with accurate secret keys, signifying authorized users, can decrypt and reveal authentic labels.
	Importantly, the proposed framework not only facilitates the protected model to multiple authorized users without requiring repetitive training of the original ML model with IP protection methods but also maintains the model's performance without compromising its accuracy. 
	Compared to existing methods like watermark-based, trigger-based, and passport-based approaches, \textit{EncryIP} demonstrates superior effectiveness in both training protected models and efficiently detecting the unauthorized spread of ML models.
\end{abstract}

\section{Introduction}

Recent years have witnessed a surge in interest regarding intellectual property (IP) protection for ML models, with attention spanning both academia and industry \cite{DBLP:conf/glvlsi/Xue0L21,DBLP:journals/corr/abs-1811-03713,TAUHID2023100114}. The central theme of current approaches revolves around embedding distinctive information within the ML model or its training data. This empowers users to verify ownership of the ML model by confirming this embedded information.
Broadly, these methods can be categorized into three types:
(1) \textbf{Watermark-based method:} It incorporates designated watermarks into model parameters, strategically employing regularization terms for embedding \cite{DBLP:journals/corr/abs-2009-12153,DBLP:journals/pami/ZhangCLZFHY22}; (2) \textbf{Trigger-based method:} This method assigns a specific value to a trigger set during model training. The model outputs this value when the same trigger set is used for verification \cite{DBLP:conf/ccs/ZhangGJWSHM18}; and (3) \textbf{Passport-based method:} By introducing a new passport layer into the original deep neural network model, this technique uses a set of data as the passport for ownership verification.
The performance of the original task will significantly deteriorate if a wrong passport is used \cite{9454280}.

While the mentioned methods can indeed create protected ML models, they remain inadequate in addressing the issue of unauthorized model distribution—when the model is used by unintended users. For instance, envision an AI company developing and licensing an ML model to multiple users. The company's aim is to restrict the model's use to authorized users solely. However, practical scenarios often witness authorized users sharing the model with unauthorized counterparts, bypassing the company's consent. Moreover, the risk persists of the model being stolen and employed by unauthorized entities.

To address this challenge, a prevailing strategy entails allocating different versions of same model to various authorized users. By overseeing the utilization of these model variations, the model creator can discern whether a specific version has been disseminated to unintended users.
However, directly implementing this strategy necessitates creating separate protected model iterations for each user. This approach entails multiple rounds of model training using existing IP protection techniques. Unfortunately, this method becomes impractical in real-world scenarios due to several reasons:

$\bullet$ \textbf{Large-Scale Models.} 
Training large-scale models such as GPT-4 \cite{DBLP:journals/corr/abs-2303-08774} or DALL-E \cite{DBLP:conf/icml/RameshPGGVRCS21} takes substantial time, often spanning days or months. Requiring multiple training for these models would incur significant costs in terms of computing resources, time, and manpower.

$\bullet$ \textbf{Restricted Usage.} In certain scenarios, like military or sensitive data, the utilization of data might be limited due to confidentiality agreements or proprietary concerns. In such cases, employing a strategy involving multiple rounds of training is impractical.

Based on the above demands of practice, we present  \textit{EncryIP}, an \underline{Encry}ption-based framework for \underline{I}ntellectual property \underline{P}rotection of machine learning model. This approach generates multiple safeguarded model versions through a single training iteration.
\textit{EncryIP} considers a special kind of public-key encryption scheme, focusing on the dataset, particularly within the label space. 
Our framework unveils a novel encryption strategy, bridging the relationship between encryption schemes and machine learning algorithms. Empirical investigations encompass various ML models, validating the efficacy and efficiency of our approach.

The contributions of the paper are summarized as follows:

$\bullet$ We introduce a novel encryption-based framework called \textit{EncryIP}, to address model intellectual property protection. Compared with existing solutions, \textit{EncryIP} generates multiple protected versions of the same model through a single training process, eliminating the need for repetitive training. 
Furthermore, the protected model produced by \textit{EncryIP} utilizes randomized labels, preventing unauthorized users without accurate secret keys from accessing true labels. Our approach satisfies IP protection requirements while surpassing the efficiency of existing methods.

$\bullet$ \textit{EncryIP} represents a data-dependent strategy utilizing a public-key encryption scheme that interacts directly with the dataset. It avoids extensive model structural modifications or additional trigger datasets. It creates a degree of independence between the learning algorithm and the encryption algorithm, which makes it easier to operate on different structures of ML models. \textit{EncryIP} provides a practical and versatile solution for protecting the intellectual property of machine learning models.

$\bullet$ We evaluate \textit{EncryIP} on a great diversity of scenarios, including different model structures and data sets. The evaluation results provide valuable insights into the effectiveness and efficiency of \textit{EncryIP} in real-world settings, demonstrating its ability to handle different model architectures and datasets with satisfactory outcomes.

\section{Related Work}


One direction is the \textbf{watermark-based method} \cite{DBLP:conf/mir/UchidaNSS17,DBLP:conf/uss/AdiBCPK18,DBLP:journals/nca/MerrerPT20,DBLP:conf/asplos/RouhaniCK19,DBLP:conf/apsipa/KuribayashiTF20,DBLP:conf/ksem/FengZ20,DBLP:conf/mir/ChenRFZK19,DBLP:conf/iccad/GuoP18,DBLP:journals/corr/abs-1904-00344,DBLP:conf/apsipa/KuribayashiTF20}. The basic idea is to design a specific embedding mechanism and utilize this mechanism in the training process. The specific embedding information can be viewed as one kind of watermark. User can verify the right of ownership by watermark. For instance,
\citeauthor{DBLP:conf/mir/UchidaNSS17}~\cite{DBLP:conf/mir/UchidaNSS17}
proposed a general framework for embedding a watermark in model parameters by using a parameter regularizer. This regularizer can be used to embed a $T$-bit vector into model parameters, and all watermarks can be correctly detected by a simple linear transformation. For real applications, \citeauthor{DBLP:conf/iccad/GuoP18}~\cite{DBLP:conf/iccad/GuoP18} proposed a watermarking framework enabling DNN watermarking on embedded devices.

The second is the \textbf{trigger-based method} \cite{DBLP:conf/ccs/ZhangGJWSHM18,DBLP:conf/iclr/LukasZK21,DBLP:conf/asiaccs/CaoJG21,DBLP:conf/gcce/PyoneMK20,app11030999,DBLP:journals/corr/abs-1712-05526,DBLP:conf/pakdd/ZhongZ0G020,DBLP:conf/asiaccs/CaoJG21,DBLP:conf/aaai/ZhangCLFZZCY20}. Generally, the algorithm employs a set of instances as a trigger set and embeds this trigger set information into ML model in the training procedure. The verification process is that model can output a specific result when this trigger set is treated as the input. 
For example, \citeauthor{DBLP:conf/iclr/LukasZK21}~\cite{DBLP:conf/iclr/LukasZK21} proposed a fingerprinting method for deep neural network classifiers that  extracts this trigger set from the original model, and this trigger set should have the same classification results with a copied model. \citeauthor{DBLP:conf/gcce/PyoneMK20}~\cite{DBLP:conf/gcce/PyoneMK20} proposed a model protection method by using block-wise pixel shuffling with a secret key as a preprocessing technique to input images and training with such preprocessed images. In \cite{app11030999}, a multitask learning IP protection method was proposed by learning the original classification task and the watermarking task together.

In recent years, a new way to rethink this problem has been proposed by introducing a novel \textbf{passport-based method} \cite{9454280,DBLP:conf/nips/FanNC19,DBLP:conf/nips/Zhang00Z0Y20}. The main motivation of embedding digital passports is to design and train DNN models in a way such that their inference performances of the original task (i.e.,  classification accuracy) will be significantly deteriorated due to the forged signatures. \citeauthor{DBLP:conf/nips/Zhang00Z0Y20}~\cite{DBLP:conf/nips/Zhang00Z0Y20} proposed a new passport-aware normalization formulation which builds the relationship between the model performance and the passport correctness. One advantage of the passport-based method is that it is robust to network modifications and resilient to ambiguity attacks simultaneously.

While current methods can generate protected models, they remain inadequate in addressing unauthorized model spread. For instance, high training costs make the watermark-based approach impractical, as it requires multiple training iterations for generating distinct versions. Although the trigger-based method achieves this with one training, its output is a true prediction, lacking security. Similarly, the passport-based method necessitates multiple passport groups, requiring complex training, and intricate modifications to the original model structure. In contrast, our \textit{EncryIP} framework efficiently tackles unauthorized spread, providing effective protection without the drawbacks posed by other methods.

\section{Preliminaries}
\noindent{\bf The IP protection}. Let training data set $D = \{(x_i,y_i)\}^N_{i=1}$, where $x_i \in \mathbb{R}^d$ is a training instance and $y_i \in \mathbb{Y} =\{1, 2,\ldots,z\}$ is the associated class label. A machine learning model $\mathcal{M}$ is learned from $D$ through an algorithm architecture $\mathcal{A}$. Following the definition in \cite{9454280}, the two main processes in the IP protection task are as follows:
\begin{enumerate}
	\item Learning process $L(D, \mathcal{A}, \mathcal{A}_{\text{IP}}) = [\mathcal{M}]$, is a model training process that takes training data $D$ and an IP protection method $\mathcal{A}_{\text{IP}}$ as inputs, and outputs a model $[\mathcal{M}]$. $[\cdot]$ is indicated as having the capacity of IP protection. 
	\item Deploying process $P([\mathcal{M}],x,\mathcal{A}^{-1}_{\text{IP}}) = y $, is that $[\mathcal{M}]$ outputs a results $y$ when input $x$. Note that the output $y$ is correct if and only if the user utilizes the correct inverse IP protection method $\mathcal{A}^{-1}_{\text{IP}}$, otherwise, a confused result will be generated.
\end{enumerate}

In the \textbf{watermark-based method}, $\mathcal{A}_{\text{IP}}$ involves implanting distinct information into an ML model, often through specialized regularizers in the loss function. The \textbf{trigger-based method} relies on $\mathcal{A}_{\text{IP}}$ to introduce a specific dataset that can be learned during training. During verification, model $[\mathcal{M}]$ responds with the designated information upon receiving this unique dataset as input. In the \textbf{passport-based method}, $\mathcal{A}_{\text{IP}}$ encompasses a novel structure integrated into the original model $\mathcal{M}$. The protected model $[\mathcal{M}]$ furnishes accurate outputs only upon receiving predetermined inputs.

In this paper, we elaborate on the concept of unauthorized model spread within the context of IP protection as follows:

\begin{definition}{(Unauthorized spread of model).}
	In the IP protection task, the output of a learning process $[\mathcal{M}]$ normally is authorized to a set of users $O=\{o_1,o_2,\ldots,o_J\}$. The unauthorized spread of model is that $[\mathcal{M}]$ is used by a user $o \notin O$.
\end{definition}

To meet this problem, we introduce a verification process in the IP protection task: 
\begin{itemize}[leftmargin=2em]
	
	\item Verification process $V([\mathcal{M}],o)= \{\text{True, False}\}$, is an authorization verification that verifies if a model user $o$ belongs to an authorized model user set.
\end{itemize}

\noindent{\bf PKE with multiple secret keys.} In this paper, \textit{EncryIP} incorporates a distinct form of public-key encryption (PKE) scheme \cite{KL20}, characterized by the following requisites:
i) the existence of multiple secret keys corresponding to a public key (i.e., each one of these secret keys can be used for decryption), and ii) the existence of some ill-formed ciphertexts, such that decrypting them with different secret keys (corresponding to the same public key) will lead to different messages. 

To be specific,  in this paper, a PKE scheme with the above properties  consists of the following  probabilistic algorithms.
\begin{itemize}
	\item \underline{$\textsf{\upshape{Gen}}$:} This is the key generation algorithm. It takes 
	a number $P\in\mathbb{N}$ as input,  and outputs a public key $pk$ and $P$ secret keys  $\{sk_j\}_{j=1}^{P}$. 
	\item \underline{$\textsf{\upshape{Enc}}$:} This is the encryption algorithm, taking $pk$ and a message $m$ as input, and outputting  a ciphertext $c$. 
	\item \underline{$\textsf{\upshape{Dec}}$:} This is the decryption  algorithm, taking $sk$ and a ciphertext $c$ as input, and outputting a message $m$ or $\bot$, which indicates that $c$ is invalid. 
	\item \underline{$\textsf{\upshape{Fake}}$:} This is the fake  encryption algorithm, taking $pk$ as input, and outputting  an ill-formed ciphertext $c$. 
\end{itemize}
For correctness, we require that for any valid message $m$, and any  $(pk,\{sk_j\}_{j=1}^{P})\leftarrow\textsf{\upshape{Gen}}(P)$: 
\begin{enumerate}[leftmargin=2em]
	\item[(i)] for  any $j\in\{1,\cdots,P\}$,  $\textsf{\upshape{Dec}}(sk_j, \textsf{\upshape{Enc}}(pk, m))=m$.
	\item[(ii)] for any $c\leftarrow\textsf{\upshape{Fake}}(pk)$ and any distinct $j_1,j_2\in\{1,\cdots,$ $P\}$, $\textsf{\upshape{Dec}}(sk_{j_1}, c)\neq\textsf{\upshape{Dec}}(sk_{j_2}, c)$. 
\end{enumerate}

\section{The Proposed Framework}




\subsection{\textit{EncryIP}: An overview}

The proposed \textit{EncryIP} framework integrates a distinctive form of public-key encryption (PKE) scheme directly within a learning algorithm, primarily within the dataset's label space. \textit{EncryIP} aims to achieve the following objectives:
(1) It generates an interpretable output exclusively when the model user employs the correct secret key.
(2) Different users possess distinct secret keys, enabling \textit{EncryIP} to identify the unauthorized user during model misuse.
(3) Integrating an encryption scheme does not significantly compromise the predictive performance of the ML model.




To realize the aforementioned objectives, \textit{EncryIP} employs a three-step process:









\begin{itemize}
	\item \textsf{Learning Process.} A learning process $L(D,\mathcal{A}, \textsf{PKE}) =  ([\mathcal{M}],$ $\{sk_j\}_{j=1}^P)$ first processes training data $D$ by a PKE scheme \textsf{PKE}, and then executes a learning algorithm $\mathcal{A}$ to output the model $[\mathcal{M}]$ and a set of secret keys $\{sk_j\}_{j=1}^P$.

	


	
	

	
	\item \textsf{Deploying Process.} A deploying process $P(x, [\mathcal{M}],$ $sk_j)= y$ is that the protected model $[\mathcal{M}]$ outputs the prediction result $y$ when $x$ is as its input. 
	
	\textit{Remark}: In general, the output of model $[\mathcal{M}]$ is correct if and only if the user uses correct $sk_j$ to decrypt it, otherwise a confused result will be output.

	
	\item \textsf{Verification Process.} An verification process $V([\mathcal{M}],$ $sk_j)= \{\text{True}, \text{False}\}$ is to evaluate if $[\mathcal{M}]$ belongs to its owner $o_j$. 
	
\end{itemize}

The crucial aspect of \textit{EncryIP} lies in the integration of an encryption scheme with a learning algorithm $\mathcal{A}$. In this paper, we propose a data-dependent IP protection method where the encryption scheme primarily focuses on the data, particularly in the label space. It can be described by
\begin{equation*}
	\small
	\min_{[\mathcal{M}]} \ \ \sum_{i=1}^{N} (\Pr [\psi^{-1} ([\mathcal{M}](x_i))=y_i)]  - \Pr [\mathcal{M}(x_i)=y_i])
\end{equation*}
where $\psi$ is an encryption function on a dataset $D$ by $\psi(D) = (x_i, \psi_i(y_i))$, and $\psi^{-1}$ is a decryption function, $D= \psi^{-1} ( \psi(D) )$. $\mathcal{M}$ is a model learned from $D$, e.g., $\mathcal{M} \leftarrow \mathcal{A}(D), y \leftarrow \mathcal{M}(x)$ and $[\mathcal{M}] \leftarrow \mathcal{A}(\psi(D)), [y] \leftarrow [\mathcal{M}](x)$.





\subsection{\textit{EncryIP}: Encryption}\label{sec:SIP_enc}



Integrating encryption and machine learning poses a crucial challenge: ensuring accurate communication between the two processes. Encryption and decryption demand exactness, while machine learning often deals with approximations, like probabilities. This discrepancy requires thoughtful solutions to bridge the gap and make encryption and learning methods work harmoniously. 
In this paper, we introduce a novel encryption strategy that presents an effective transfer function between label space and encryption scheme.

First of all, we show the details of the encryption scheme here. As described before, 
we consider a PKE with multiple secret keys corresponding to the same public key, and further require that there exist some ill-formed ciphertexts, such that decrypting them with different secret keys (corresponding to the same public key) will lead to different messages. 
Many PKE schemes, like those built using hash proof systems \cite{CS02}, adhere to these characteristics. For practical purposes, we focus on a simplified form of the Cramer-Shoup scheme \cite{CS98}, which we call $\csl$.


\begin{figure}[t]  
\scriptsize
\centering
\begin{tabular}[c]{|l|}\hline
	\underline{$\gen(P)$:} 
	\\
	
	$t\leftdollar\mathbb{Z}_q^*$, $g_2=g_1^t$, $(a_1,b_1)\leftdollar(\mathbb{Z}_q)^2$, $h=g_1^{a_1}g_2^{b_1}$ 
	\\
	
	$\{b_j\leftdollar\mathbb{Z}_q\}_{j=2}^P$ (s.t. $b_j\neq b_1$ and $b_{j_1}\neq b_{j_2}$ for $j_1\neq j_2$) 
	\\
	
	
	$\{a_j=a_1+(b_1-b_j)t\}_{j=2}^P$	
	\\
	
	$pk=(g_1,g_2,h)$, $\{sk_j=(a_j,b_j)\}_{j=1}^P$	
	\\
	
	Return $(pk,\{sk_j\}_{j=1}^P)$	
	\\
	
	\hline
	
	\underline{$\enc(pk,m)$:} \\
	$r\leftdollar\mathbb{Z}_q$, $u_1=g_1^r$, $u_2=g_2^r$, $u_3=h^rm$ \\
	Return $c=(u_1,u_2,u_3)$\\
	\hline
	
	\underline{$\dec(sk,c)$:} \\
	Parse $sk=(a,b)$ and $c=(u_1,u_2,u_3)$\\
	Return $m=\frac{u_3}{u_1^au_2^b}$  \\
	\hline
	
	\underline{$\fake(pk)$:} \\
	$(r_1,r_2)\leftdollar(\mathbb{Z}_q)^2$ (s.t. $r_1\neq r_2$), 
	$u_1=g_1^{r_1}$, $u_2=g_2^{r_2}$, $u_3\leftdollar\mathbb{Z}^*_q$  \\
	Return $c=(u_1,u_2,u_3)$\\
	\hline

\end{tabular}
\caption{PKE scheme $\csl=(\gen,\enc,\dec,\fake)$.}
\label{PKE_scheme}
\end{figure}

	
	
	
	
	
	
	

Let $\mathbb{G}_q$ be a cyclic group  of some prime order $q$, $g_1$ be a generator of $\mathbb{G}_q$, and $\mathbb{Z}_q^*=\mathbb{Z}_q\setminus\{0\}$. We present a PKE scheme  $\csl=(\gen,\enc,\dec,\fake)$, with message space $\mathbb{G}_q$, as shown in Figure \ref{PKE_scheme}. Note that here for a finite set $S$, we write  $s\leftdollar S$ for sampling $s$ uniformly random from $S$. Due to space limitations, the correctness analysis is shown in Appendix.

For security of $\csl$, we present the following theorem. Due to space limitations, the definition of the DDH assumption and that of IND-CPA security are recalled in Appendix, and the proof of Theorem \ref{thm:CPA} is given in Appendix.
\begin{theorem}\label{thm:CPA}
If the DDH assumption holds for $\mathbb{G}_q$, $\csl$ is IND-CPA secure. 
\end{theorem}

Based on $\csl$  in Figure \ref{PKE_scheme}, we present our proposed encryption strategy, which encompasses three key steps:



\textbf{1. Generate Public/Secret Keys:} The process begins by utilizing the algorithm $\gen$ within $\csl$ to create public and secret key pairs.


\textbf{2. Encrypt Labels:} Working with the training set $D$ and the algorithm $\enc$, the process involves processing the training set $D$ as follows:
\begin{equation}
\small
c \leftarrow \enc (pk,y),
\label{cip}
\end{equation}
where ciphertext $c$ is a three-tuple, i.e.,  $(u_1,u_2,u_3)$. 



It is important to note that the outcome of $\enc$ is generally a random value. In our paper, we establish the first encryption result as the ciphertext for each label, and it remains unchanged subsequently.
However, in reality, the outcome of $\enc(pk,y)$ can be viewed as a randomly sampled value from a distribution denoted as $\textsf{Dist}_{y}^{\text{ct}} =
\{r\leftdollar\mathbb{Z}_q:(g_1^r, g_2^r, h^ry)\}$. This distribution is characterized by the following attributes:
\begin{itemize}
\item The distribution can be represented by any element  sampled from it. In other words, given an element $c'$ sampled from $\textsf{Dist}_{y}^{\text{ct}}$, $\textsf{Dist}_{y}^{\text{ct}}$ can be represented with $c'$.
\end{itemize}
That is because for any element $c'$ sampled from $\textsf{Dist}_{y}^{\text{ct}}$, $c'$ can be written as $c'=(u'_1,u'_2,u'_3)=(g_1^{r'}, g_2^{r'}, h^{r'}y)$. So $\textsf{Dist}_{y}^{\text{ct}}$ can be written as 
\begin{small}
\[\textsf{Dist}_{y}^{\text{ct}}=\{r\leftdollar\mathbb{Z}_q:(g_1^{r}u'_1,g_2^{r}u'_2,h^ru'_3)\}.\]
\end{small}
Hence, (1) the ciphertext $c\leftarrow\enc(pk,y)$ can be viewed as a representation of distribution; and (2) the following learning process can also be viewed as learning on distribution corresponding to labels, rather than a strict label value.

Additionally, we define a sampling algorithm $\textsf{SampDist}$, which takes  a cipehrtext $c'=(u'_1,u'_2,u'_3)$ sampled from $\textsf{Dist}_{y}^{\text{ct}}$ as input, and outputs a new ciphertext sampled from   $\textsf{Dist}_{y}^{\text{ct}}$. Formally, we write  $c\leftarrow\textsf{SampDist}(c')$ for this process. The detailed description of  $\textsf{SampDist}$ for the encryption algorithm $\enc$
of $\csl$ is as follows. 
\begin{description}\small
\item{\underline{$\textsf{SampDist}(c'=(u'_1,u'_2,u'_3))$:}}\\
$r\leftdollar\mathbb{Z}_q$; $~u_1=g_1^ru'_1$; $~u_2=g_2^ru'_2$; $~u_3=h^ru'_3$\\
Return $c=(u_1,u_2,u_3)$
\end{description}



\textbf{3. Transfer ciphertexts to ``confused'' labels}. As the structure of $c$ is not a normal label structure, a learning algorithm $\mathcal{A}$ can not directly use it to train. 
In the ensuing section, we present a transfer function that establishes a connection between a conventional label structure and the ciphertext structure.
We refer to every true label $y$ as a ``readable'' label. For any label $[y]$ generated from a transfer function, we refer to it as a ``confused'' label. 
\begin{definition}{(Transfer Function).}
Denote  $\Phi$ as the \emph{transfer function} such that for each ciphertext $c$ in the ciphertext structure,  $\Phi(c) = [y]$. Denote $\Phi^{-1}$ as  the inverse function of $\Phi$ satisfying $\Phi^{-1}(\Phi(c))=c$  for all ciphertexts.
\label{define:PF}
\end{definition}


In $\csl$, the function $\Phi$ can be viewed as an indicator function $\mathbb{I}$ which outputs a $q$-dimension\footnote{$q$ is a parameter related to security of the encryption scheme. In our implementation, for simplicity, we set $q$ to the smallest prime number greater than or equal to the number of classes.} all-zero vector except the position $c$ is $1$, 
\begin{equation}
\small
T \in \{0,1\}^q \leftarrow \  \Phi(c) = \mathbb{I}(c)
\end{equation}
\noindent Specifically, as described in Eqn. (\ref{cip}), the ciphertext is a three-tuple $(u_1,u_2,u_3)$. We take $u_1,u_2$ and $u_3$ separately as the input in $\Phi$ and get $T_1$, $T_2$ and $T_3$. Then, we concatenate them as $[y]$, i.e., $[y]=\{T_1 T_2 T_3\}$. Note that $[y]$ can be used in a normal learning process, but $[y]$ is a confused label that is totally different from $y$.


\begin{figure}[t] 
\centering 
\includegraphics[width=0.4\textwidth]{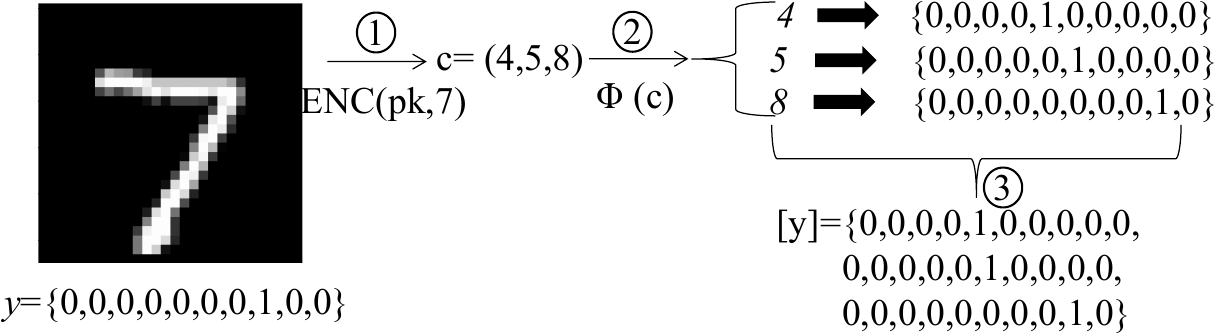} 
\caption{A case study on encryption process. (1) The index of label is encrypted by encryption algorithm \enc; (2) A ciphertext is transferred to label structure by transfer function $\Phi$; (3) A confused label is achieved by concatenation.}
\label{Fig:casestduy} 
\end{figure}


After that, we obtain a new dataset $[D]=\{x_i,[y_i]\}_{i=1}^N$. A case study is shown in Figure \ref{Fig:casestduy}.

\vspace{1mm}
\textit{Remark}: 
(1) It is evident that by using alternative PKE schemes constructed from various hash proof systems (e.g., \cite{CS02,HK07,NS09}) as foundational components, the ciphertext can be configured as a $k$-tuple (i.e., $c = (u_1,u_1,\ldots,u_k)$). Generally, a higher value of $k$ enhances the security of the PKE but results in reduced efficiency. For practicality and clarity, we adopt a simplified version with $k=3$ in this paper. (2) It's worth noting that $[y]$ can also be calculated using different methods, such as a reduction technique, e.g., $[y]={V_1 + V_2 + V_3}$.

\begin{algorithm}[t] \small
\SetKwInOut{Input}{Input}\SetKwInOut{Output}{Output}
\caption{Learning Process}
\label{TP}
	\Input{$D = \{(x_i,y_i)\}^N_{i=1}$ - input data. $\mathcal{A}$ - algorithm architecture. $\csl$ - encryption scheme.}
		\Output {$[\mathcal{M}]$ - model, $\{sk_j\}_{j=1}^P$  - $P$ secret keys.}
		$(pk,\{sk_j\}_{j=1}^{P}) \leftarrow$ $\gen(P)$ \;
		\For { $ i=1,\ldots,N $}{
			$c_i \leftarrow $ $\enc(pk,y_i)$ \;
			$[y_i] \leftarrow \Phi(c_i)$ \;} 
		$[\mathcal{M}] \leftarrow \mathcal{A}(\{(x_i,[y_i])\}^N_{i=1})$ \;
		\Return $[\mathcal{M}], \{sk_j\}_{j=1}^P$.
	
\end{algorithm}

\subsection{\textit{EncryIP}: Learning /Deploying}

\textbf{Learning.} Once the data is encrypted, the learning process is executed on the encrypted dataset $[D]$. The comprehensive learning process of \textit{EncryIP} is delineated in Algorithm \ref{TP}. Notably, the training procedure for the ML model remains unaltered from the original model, as illustrated in line 6 of Algorithm \ref{TP}. The sole modification lies in replacing the original loss function with a soft-label loss within the learning algorithm $\mathcal{A}$. In Sec.  \ref{effectivenessofSIP}, we demonstrate that this adjustment does not hinder the convergence of the learning process.




\noindent \textbf{Deploying.} The deployment process is formally outlined in Algorithm \ref{DP}. In our framework, \textit{EncryIP} produces a random prediction (confused label), thereby preventing attackers from inferring the true label through analysis of the confused label.

The deploying process unfolds as follows:

(1) Initially, an inverse function is utilized to convert confused labels into ciphertexts. For instance, assuming $[y]$ is an output vector denoting label probabilities, the inverse function $\Phi^{-1}$ operates as:
\begin{equation}
	\small
	c \leftarrow \Phi^{-1}([y]) = I([y])
\end{equation}
Here, $I()$ returns the position index of the top $k$ values in input $[y]$. For instance, when $k=3$, $I()$ generates a ciphertext $c$ from the position index of the top 3 values in prediction $[y]$, i.e., ${I_1,I_2,I_3}$ is treated as ciphertext $c$.


(2) As previously discussed, $c$ can symbolize a distribution $\textsf{Dist}{y}^{\text{ct}}$ linked to a certain $y$. In line 3 of Algorithm \ref{DP}, a random ciphertext $c_d$ is derived by sampling from $\textsf{Dist}_{y}^{\text{ct}}$.

(3) Ultimately, \textit{EncryIP} generates a random prediction (a confused label), as depicted in lines 4 to 5 of Algorithm \ref{DP}.

An authorized user holding the correct secret key $sk_j$ ($j\in{1,\cdots,P}$) can obtain the actual label:
\begin{equation}
	\small
	y \leftarrow \dec(sk_j,c_d).
	\label{dece}
\end{equation}

\begin{algorithm}[t] \small
	\SetKwInOut{Input}{Input}\SetKwInOut{Output}{Output}
	\caption{Deploying Process}
	\label{DP}
		\Input{$x$ - input dataset. $[\mathcal{M}]$ - model, $sk_j, j \in \{1, 2,\ldots,P\}$ - a secret key.}
			\Output{ $[y_d]$ - a confused label.}
			$[y] \leftarrow [\mathcal{M}](x)$ \;
			$c \leftarrow \Phi^{-1}([y])$    \;
			$c_d \leftarrow\textsf{SampDist}(c)$ \;
			$[y_d] \leftarrow \Phi(c_d)$ \;
			\Return $[y_d]$.
	\end{algorithm}
	


		
		
		
		
		
	
	\subsection{\textit{EncryIP}: Verification}
	In this section, we describe the methodology for identifying unauthorized usage of the model. We assume a scenario where a neutral arbitrator, who possesses all secret keys $S=\{{sk}_j\}_{j=1}^P$, sells the ML model to $P$ users, each possessing a distinct secret key. The verification procedure is carried out by the arbitrator. We consider the two following scenarios:
	
	
	
		$\bullet$ \textbf{Secret key available.} In this scenario, the arbitrator can know which secret key is used by a user in the deploying process.
		This is a simple and easy verification scenario that the neutral arbitrator can check if the secret key belongs to a corresponding model user by:
		\begin{small} 
			\begin{numcases}{V([\mathcal{M}],sk)=}
				j, & if $sk \in S$, $sk = sk_j$  $\nonumber$ \\
				$False$, &  if $sk \notin S$ $\nonumber$
			\end{numcases}
		\end{small}
		

		
		$\bullet$ \textbf{Secret key unavailable.} 
		In this scenario, the arbitrator lacks knowledge about which secret key a user employs during the deployment. Detecting unauthorized model usage without this information poses a challenge. To address this challenge, we leverage the property of $\csl$, where ill-formed ciphertexts yield distinct decrypted messages based on different secret keys. This property can be applied to \textit{EncryIP} as follows: 


		
		
		\begin{figure}[t] 
			\centering 
			\includegraphics[width=0.4\textwidth]{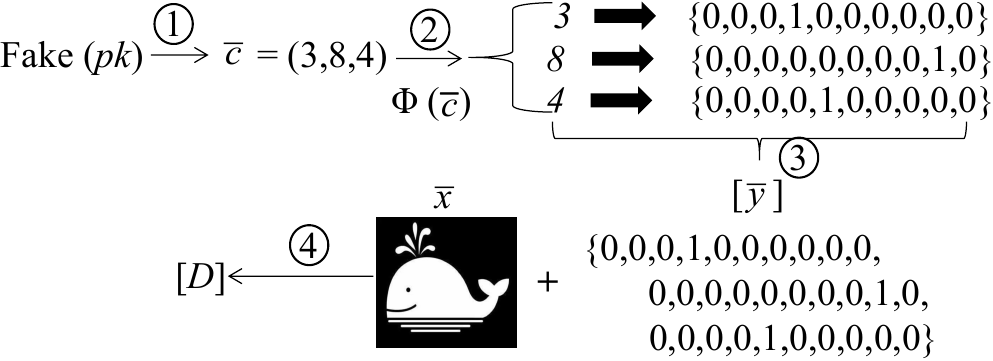} 
			\caption{A case study on verification (when the secret key is unavailable). (1) One kind of ill-formed ciphertext $\bar{c}$ is generated by algorithm \fake; (2) A ciphertext is transferred to label structure by transfer function $\Phi$; (3) A confused label is achieved by concatenation; (4) A verification instance-label pair $\{\bar{x}, [\bar{y}]\}$ is added to $[D]$.}
			\label{Fig:casestduy4} 
		\end{figure}
		
		Let $\{\bar{x}, [\bar{y}] \}$ be a verification instance-label pair. $\bar{x}$ is an arbitrary instance (e.g., an arbitrary picture), and $\bar{y}$ is generated from one kind of ill-formed ciphertext $\bar{c}$ (i.e., $[\bar{y}] \leftarrow \mathbb{I}(\bar{c})$), where $\bar{c}\leftarrow\fake(pk)$. 
		We assume $\{\bar{y}_j'\}_{j=1}^P$ is a set of results which is from decrypting the ciphertext $\bar{c}$ by using different secret keys, i.e., 
		$\bar{y}_j' \leftarrow \dec(sk_j,\bar{c})$ for $j \in \{1,\ldots,P\}$. Note that the arbitrator has this information $\{\bar{x},\{sk_j,\bar{y}'_j\}_{j=1}^P\}$. Then we add $\{\bar{x}, [\bar{y}]\}$ to $[D]$, and train the ML model $[\mathcal{M}]$. A simple case is illustrated in Figure \ref{Fig:casestduy4}.


		
		When $\bar{x}$ is utilized as input and distinct secret keys are employed to decrypt the ciphertext $\bar{c}$, $[\mathcal{M}]$ will yield varying results due to the aforementioned property. Consequently, the verification process is as follows:
		\begin{footnotesize}
			\begin{equation}\label{vf}
				V([\mathcal{M}],\bar{x})=
				\begin{cases} 
					j, & \mbox{if} \ \mbox{$~\dec(sk_j,\bar{c}) = \bar{y}'_j$}. \\
					\text{False},  &  \begin{matrix} 
						\mbox{if } \ \mbox{$~\forall~j\in\{1,\cdots,P\},$} \\
						\mbox{$~ \dec(sk_j,\bar{c})\neq\bar{y}'_j$.} 
					\end{matrix} \\
				\end{cases}
			\end{equation}
		\end{footnotesize}
		where $\bar{c}\leftarrow I([\mathcal{M}](\bar{x}))$.
		
		







\subsection{\textit{EncryIP}: Analysis}
\textbf{(1) Against removal attacks and ambiguity attacks.} Removal attacks seek to eliminate or alter IP protection measures by modifying model weights through techniques like fine-tuning or pruning. Since \textit{EncryIP} is data-dependent, its performance remains unaffected regardless of the extent of fine-tuning or pruning applied. This robustness makes \textit{EncryIP} resilient against removal attacks.


Ambiguity attacks aim to forge counterfeit secret keys (or watermark, or passport) without modifying model weights. In \textit{EncryIP}, the security of $\csl$  guarantees that it is hard to forge a secret key. More specifically, according to Theorem \ref{thm:CPA}, $\csl$ is IND-CPA secure, which implies that the probability of forging a secret key successfully is negligible (otherwise, one can trivially succeed in breaking the IND-CPA security of $\csl$, contradicting Theorem \ref{thm:CPA}).
\textbf{(2) The incremental authorized users.} 
In practice, a significant challenge for IP protection methods lies in efficiently accommodating an increasing number of authorized users. 
As we know, \textit{EncryIP} produces different model versions by generating different secret keys. This challenge can be easily addressed in \textit{EncryIP}. Specifically, assume  that $P$ secret keys $\{sk_j=(a_j,b_j)\}_{j=1}^P$ have been generated, and  now a new user is authorized. In this case, a new secret key $sk_{P+1}=(a_{P+1},b_{P+1})$ can be generated as follows: sampling $b_{P+1}\leftdollar\mathbb{Z}_q$ such that $b_{P+1}\notin\{b_1,\cdots,b_P\}$, and computing $a_{P+1}=a_1+(b_1-b_{P+1})t$. Note that, this approach eliminates the need to update existing authorized users' secret keys or retrain the ML model, serving as a future direction.

\textbf{(3) The collusion.} 
The aim of this study is to systematically define and address the issue of unauthorized model distribution within the context of model IP protection. We present a robust solution under certain initial assumptions, which offers room for further enhancement. For instance, our proposed method, \textit{EncryIP}, currently operates on the premise that authorized users do not collaborate. Exploring the substitution of the employed PKE scheme in Sec. \ref{sec:SIP_enc} with alternatives rooted in distinct computational assumptions, such as the Decisional Composite Residuosity assumption \cite{Paillier99}, could yield advancements. We regard the exploration of ways to fortify \textit{EncryIP} from diverse perspectives as a promising avenue for future research.
\section{Experiment}

\subsection{The effectiveness of \textit{EncryIP}}
\label{effectivenessofSIP}

\subsubsection{The results on Learning /Deploying.} 

We evaluate the effectiveness of \textit{EncryIP} during both the learning and deployment stages. We compare the test performance of \textit{EncryIP} and $\textit{EncryIP}_{\text{Incorrect}}$ (using an incorrect secret key) with the original model. These comparisons are performed across three machine learning model structures and three datasets. We repeat each experiment 30 times and present the results in terms of mean and standard deviation.


The prediction accuracy is employed as the evaluation metric. In \textit{EncryIP}, we set $q$ to the number of classes in each data set. The common parameters in each model structure are set by default values and the same in each method. The settings of $\textit{EncryIP}_{\text{Incorrect}}$ are the same with \textit{EncryIP}, except a fake secret key is used. 


\begin{table*}[t]
	\centering
	\scriptsize
	\caption{The results on the effectiveness (Acc.). In the parentheses, the left side indicates the percentage changes between the original model and \textit{EncryIP}, and the right side indicates the performance of the verification.
	}
	\begin{tabular}{ccccccc}
		\toprule
		Model & Method & MNIST & 
		CIFAR10 & 
		CIFAR100 &  ImageNet \\ \midrule 
		\multirow{3}*{ResNet18} & original & $0.9936\pm0.0014 $ & $0.9030\pm0.0046 $  &  $0.7322\pm0.0036 $   & $0.6231\pm0.0010 $    \\ 
		& \textit{EncryIP}  & $0.9930\pm0.0006 $  ( 0.06\%, 1.0)   &  $0.9094\pm0.0051 $  ( 0.71\%, 1.0)  & $0.7279\pm0.0030 $  (0.59\%, 0.97)&  $0.6178\pm0.0006 $  (0.85\%, 1.0)    \\  
		& $\textit{EncryIP}_{\text{Incorrect}}$  & $0.0006\pm0.0003$  & $0.0075\pm0.0053 $  & $0.0027\pm0.0008 $  & $0.0011\pm0.0005$   \\ 
		\midrule 
		
		\multirow{3}*{GoogLeNet} & original & $0.9905\pm0.0011 $  & $0.8675\pm0.0036 $  & $0.6347\pm0.0031 $  & $0.6502\pm0.0016 $   \\ 
		& \textit{EncryIP}  & $0.9879\pm0.0015 $  (0.84\%, 1.0) &    $0.8508\pm0.0101 $  (1.9\%, 1.0)   & $0.6165\pm0.0053 $  (2.86\%, 1.0)   & $0.6461\pm0.0013 $  ( 0.63\%, 1.0)     \\ 
		& $\textit{EncryIP}_{\text{Incorrect}}$  & $0.0216\pm0.0403 $  & $0.0124\pm0.0087 $  & $0.0055\pm0.0013 $  & $0.024\pm0.0102 $   \\  
		\hline 
		\multirow{3}*{AlexNet} & original & $0.9914\pm0.0006 $  & $0.8563\pm0.0022 $  & $0.5439\pm0.0049 $   & $0.5673\pm0.0020 $   \\  
		& \textit{EncryIP}  & $0.9923\pm0.0010 $   (0.09\%, 1.0)   &  $0.8457\pm0.0031 $  (1.24\%, 1.0)   & $0.5282\pm0.0040 $  (2.88\%, 0.97)  & $0.5615\pm0.0020 $   (1.02\%, 1.0)    \\  
		& $\textit{EncryIP}_{\text{Incorrect}}$  & $0.0006\pm0.0003$  & $0.0181\pm0.0048 $  & $0.0055\pm0.0012 $   & $0.0006\pm0.0025$   \\  

		\bottomrule
		
	\end{tabular}
	\label{tab:effectiveness}
\end{table*}



Table \ref{tab:effectiveness} shows that the results of \textit{EncryIP} are almost the same as the performance of the original model (results for GoogLeNet and AlexNet can be found in the Appendix). 
In the parentheses, the left side indicates the percentage changes between the original model and \textit{EncryIP}, with an approximate average of $1\%$. Conversely, $\textit{EncryIP}_{\text{Incorrect}}$ exhibits subpar performance, confirming that incorrect secret keys hinder recovery in \textit{EncryIP}. These findings underscore \textit{EncryIP}'s efficacy as an impactful protection for model outputs while maintaining strong model performance.


\begin{figure}[t]
	\centering
	\subfigure[MNIST]{\epsfig{file=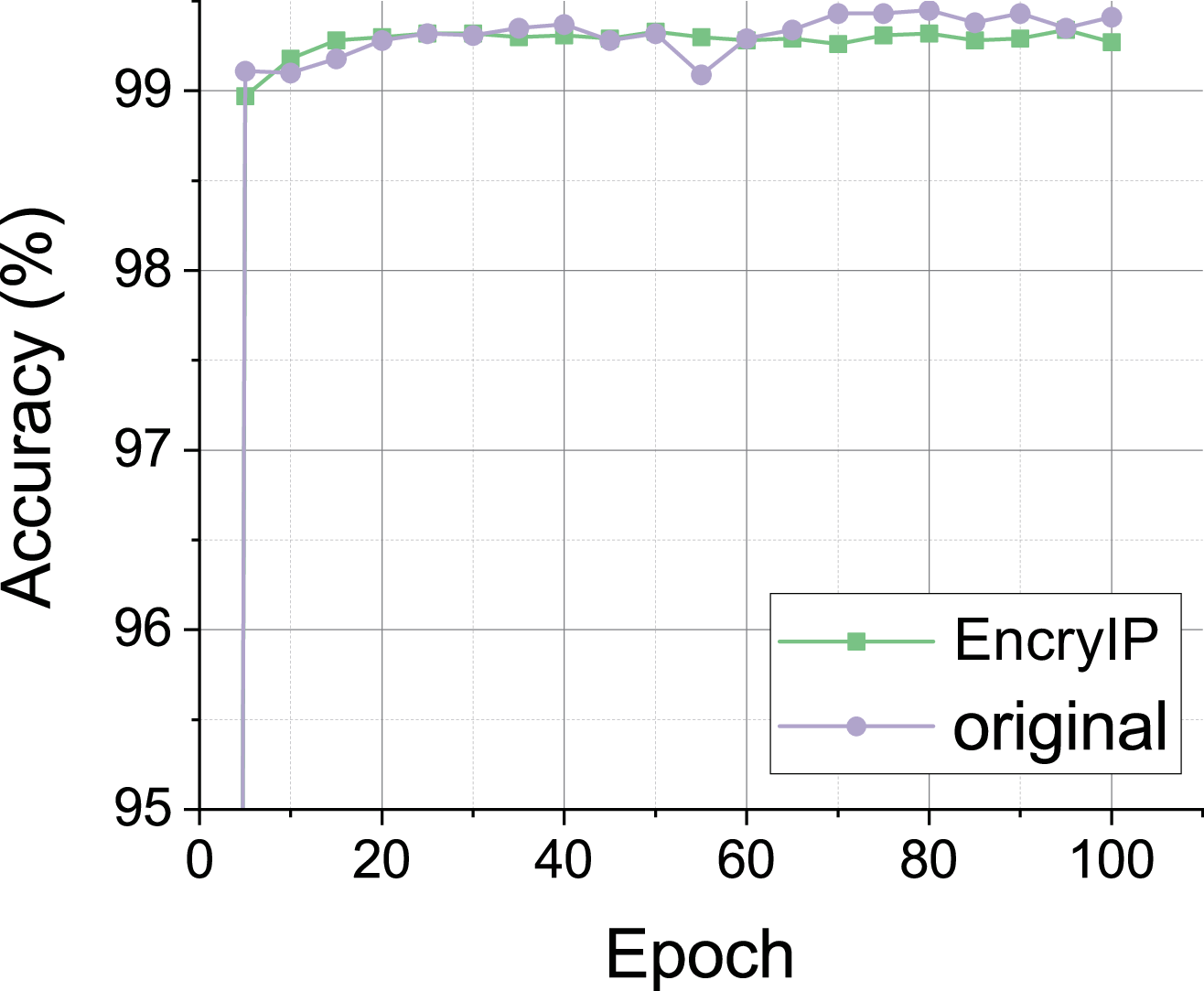,height=0.9in,width=1.05in}} 
	\subfigure[CIFAR10]{\epsfig{file=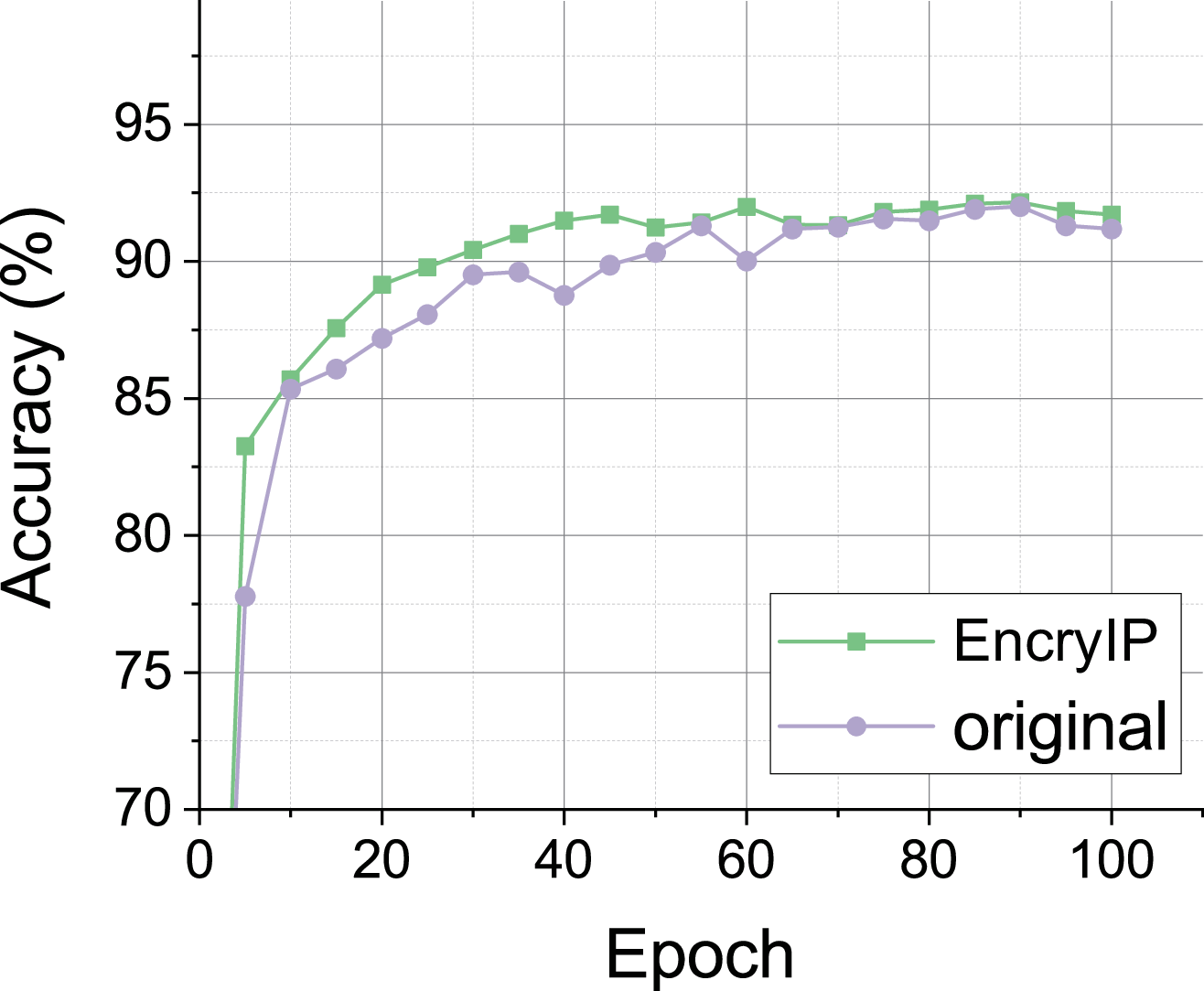,height=0.9in,width=1.05in}} 
	\subfigure[CIFAR100]{\epsfig{file=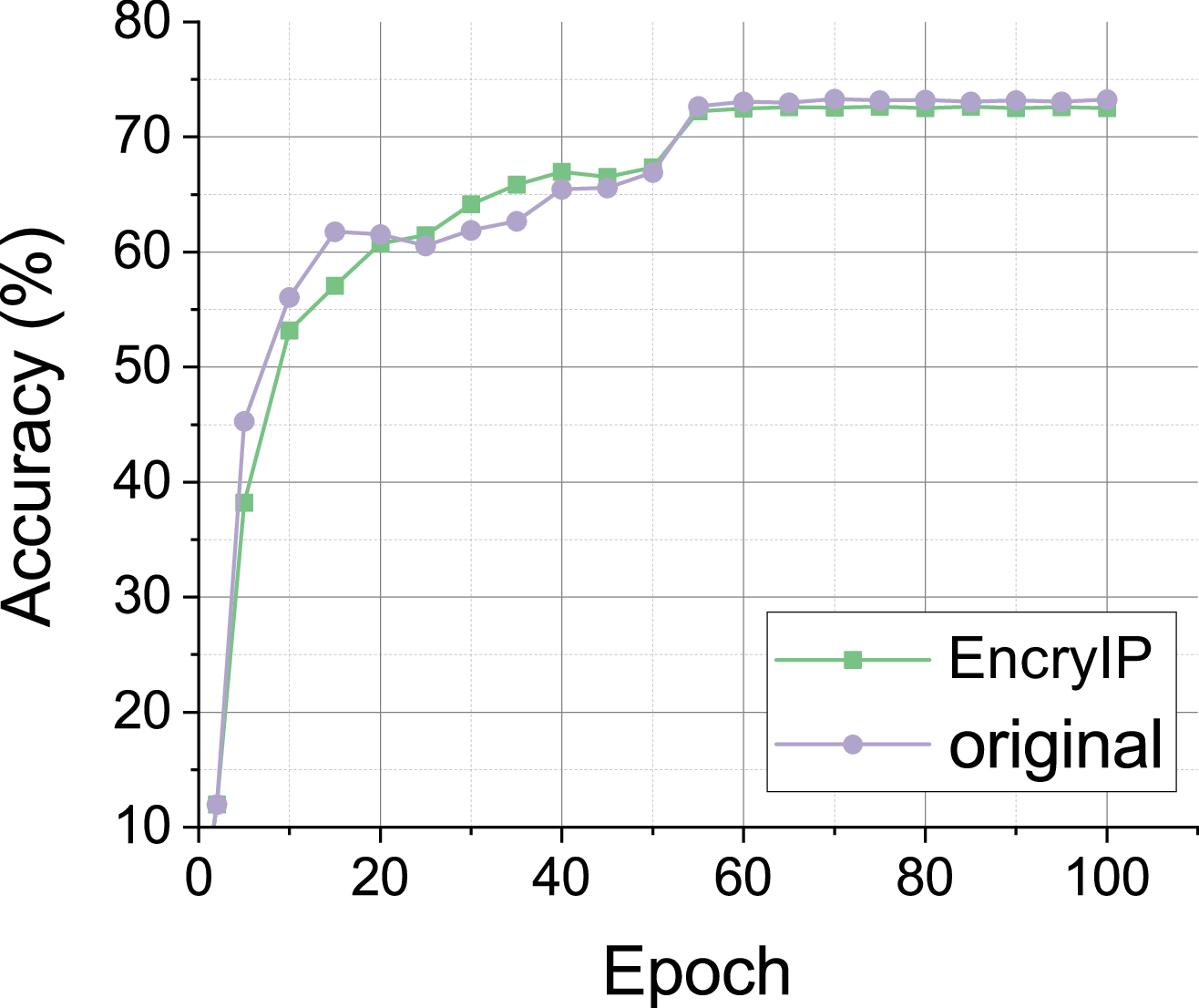,height=0.9in,width=1.05in}} 
	\caption{The training convergence for ResNet.}
	\label{fig:covergence}
\end{figure}


Figure \ref{fig:covergence} demonstrates that the incorporation of \textit{EncryIP} does not impede the convergence of the learning process. The accuracy convergence of \textit{EncryIP} aligns closely with the original model's trajectory, exhibiting nuanced differences during training. The dissimilarity between training with \textit{EncryIP} and the original approach is primarily attributed to label structure. \textit{EncryIP} introduces confused labels to the training model, potentially influencing the training process. However, as indicated in Table \ref{tab:effectiveness}, this modification has only a minimal effect on performance, highlighting the effective balance of security and performance achieved by \textit{EncryIP}.





\subsubsection{The results on verification.}\label{vtfd}
To examine the verification of $\textit{EncryIP}$ (the performance of detecting whose model is unauthorizedly used), we conducted experiments as below: The verification is under the scenario of secret key unavailable. One of the authorized user's models (e.g., $j$) is randomly selected as the model which is used by an unintended user. We check if the output of Eqn. (\ref{vf}) is equal to $j$. Note that in our experiments, the required image $\bar{x}$ in verification can be set by different types. We refer $\bar{x}$ to one kind of randomly selected image which is different from training data. The experiment is repeated 30 times, we try different types of $\bar{x}$ and different $j$ each time. In Table \ref{tab:effectiveness}, the performance of verification is shown in parentheses. The results of the verification are at nearly 100\% accuracy in different structures and different datasets.








\subsection{The efficiency of \textit{EncryIP}}

\begin{figure}[t]
	\centering
	\subfigure[CIFAR100]{\includegraphics[width=1.05in, height=0.8in]{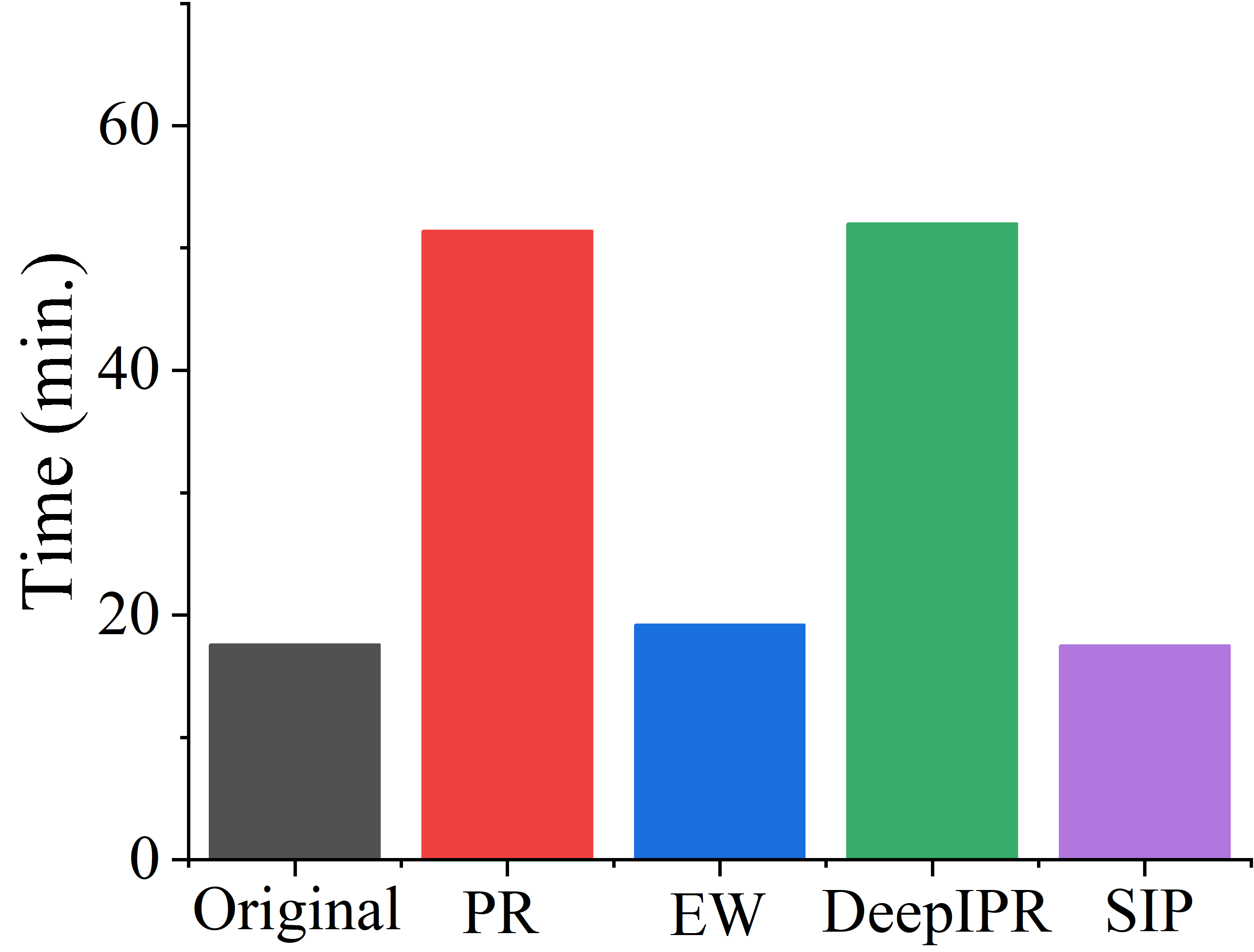}}
		\subfigure[CIFAR100]{\includegraphics[width=1.05in, height=0.8in]{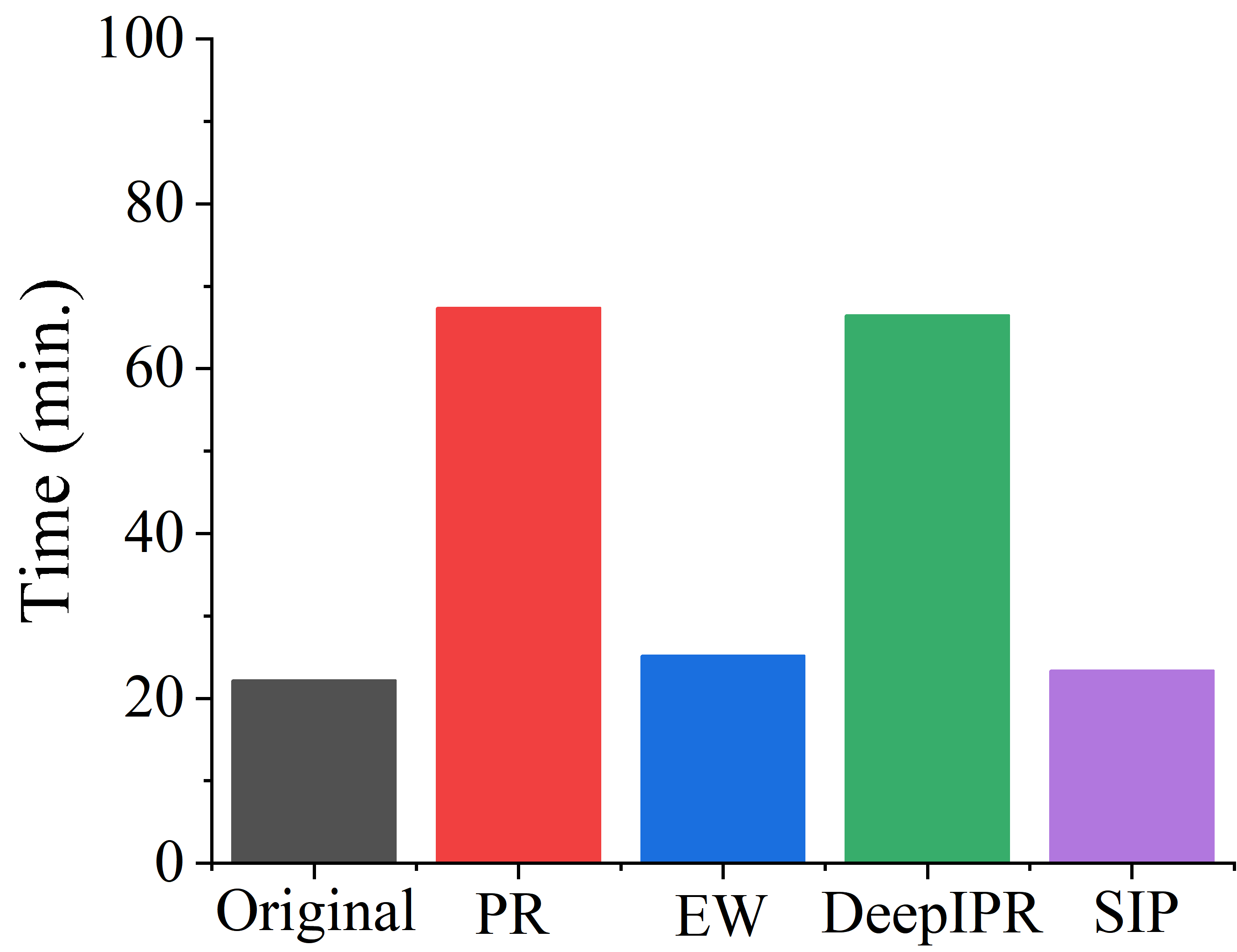}}
	\subfigure[CIFAR100]{\includegraphics[width=1.05in, height=0.8in]{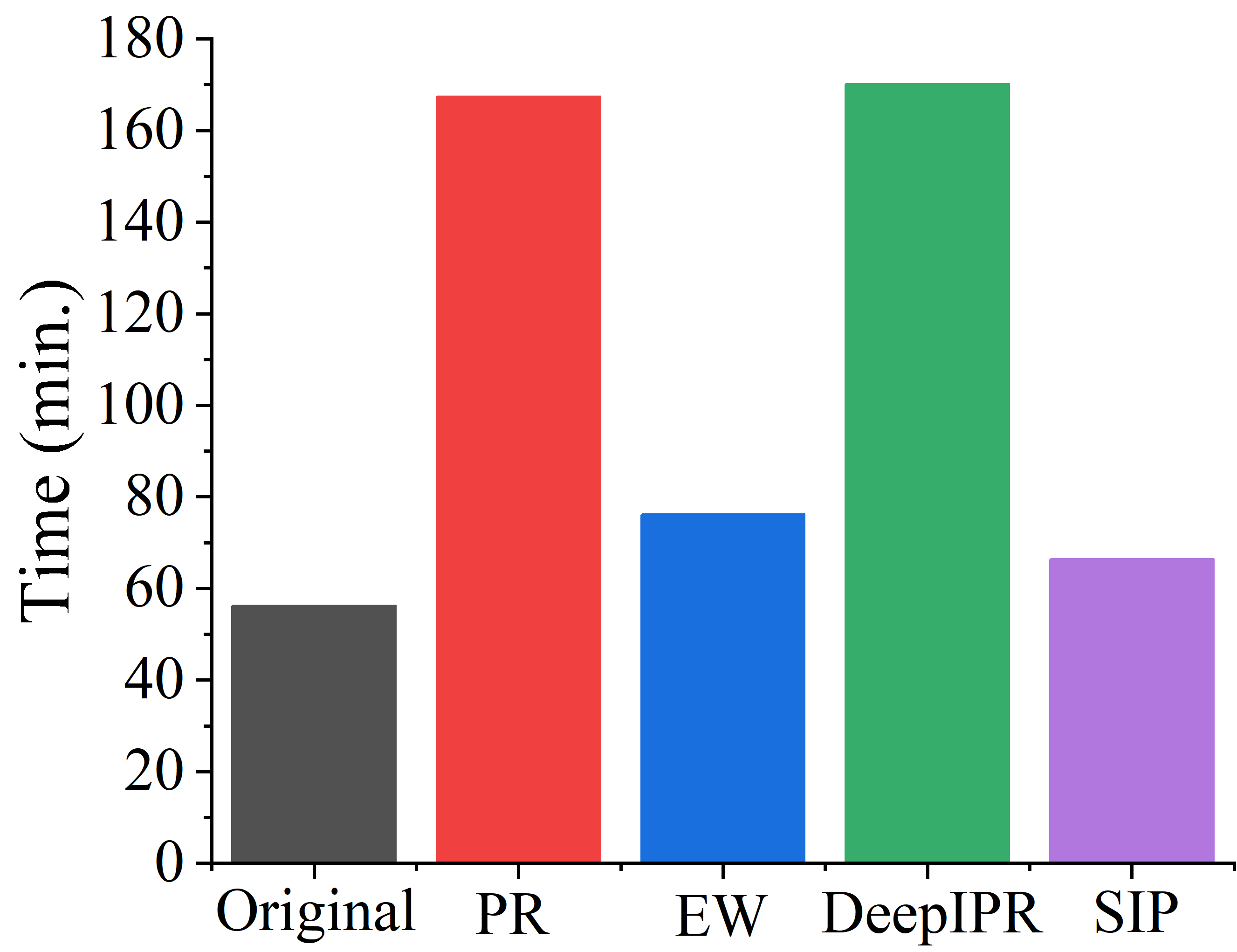}}
	\caption{The training time results of ResNet18.}
	\label{fig:time}
\end{figure}

\noindent\textbf{Settings.} We compare training times among state-of-the-art methods to highlight their efficiency in addressing the problem of the unauthorized spread of the model. 
The three typical methods from watermark-based (\textbf{PR} \cite{DBLP:conf/mir/UchidaNSS17}), trigger-based (\textbf{EW} \cite{DBLP:conf/ccs/ZhangGJWSHM18}) and passport-based method (\textbf{DeepIPR} \cite{9454280}) are chosen as the comparisons. 
The scenario involves a model producer intending to sell an ML model to three users. 
For PR, the producer must perform three separate training sessions, each yielding a distinct model version. This necessity arises from the use of distinct watermark parameters required by the embedding regularizer for generating different versions.
Similarly, DeepIPR necessitates three separate training rounds to cater to its requirement for employing three distinct passport groups to generate diverse model versions.
Although EW can train once to address this situation by incorporating three different trigger sets, it fails to meet the security requirement as its output remains interpretable.
Note that the experimental conditions are the same in each group, e.g., the number of epochs. The outcomes for ResNet18 are depicted in Figure \ref{fig:time}, while comprehensive results are available in Appendix.








\noindent\textbf{Results.} The results show that \textit{EncryIP} is more efficient than the other methods (PR and DeepIPR), and its training time is almost close to the original time (the ``Original''  in Figure \ref{fig:time}). EW performs similar results with \textit{EncryIP} because it just needs one time of training to meet this scenario. But, this approach does not provide sufficient security for the resulting predictions, as its output is a readable prediction. In contrast, our proposed method prioritizes the output of confused labels, which offers a more protected solution than EW.

\begin{figure}[t]
	\centering
		\subfigure[MNIST]{\includegraphics[width=1.05in, height=0.9in]{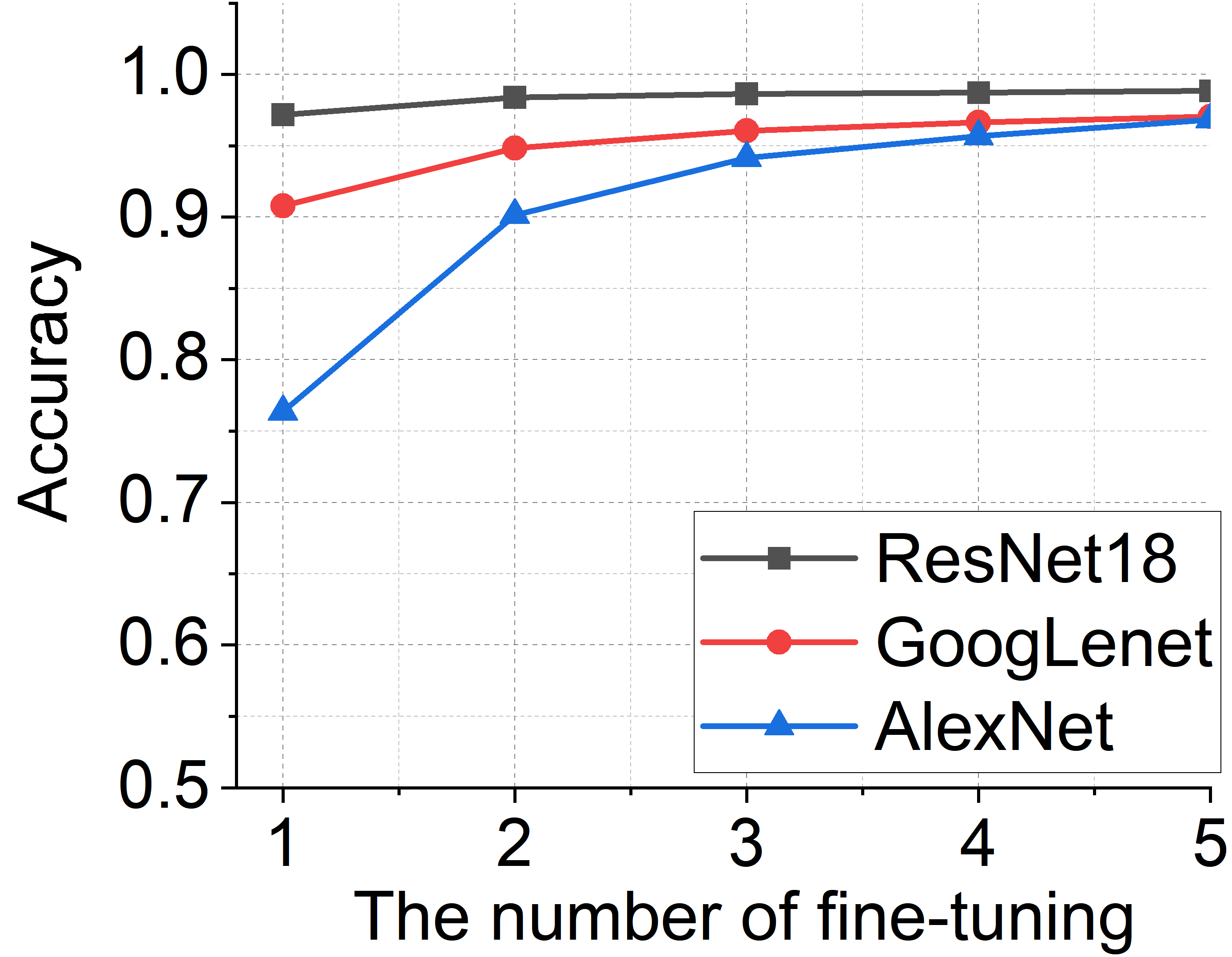}}
			\subfigure[CIFAR10]{\includegraphics[width=1.05in, height=0.9in]{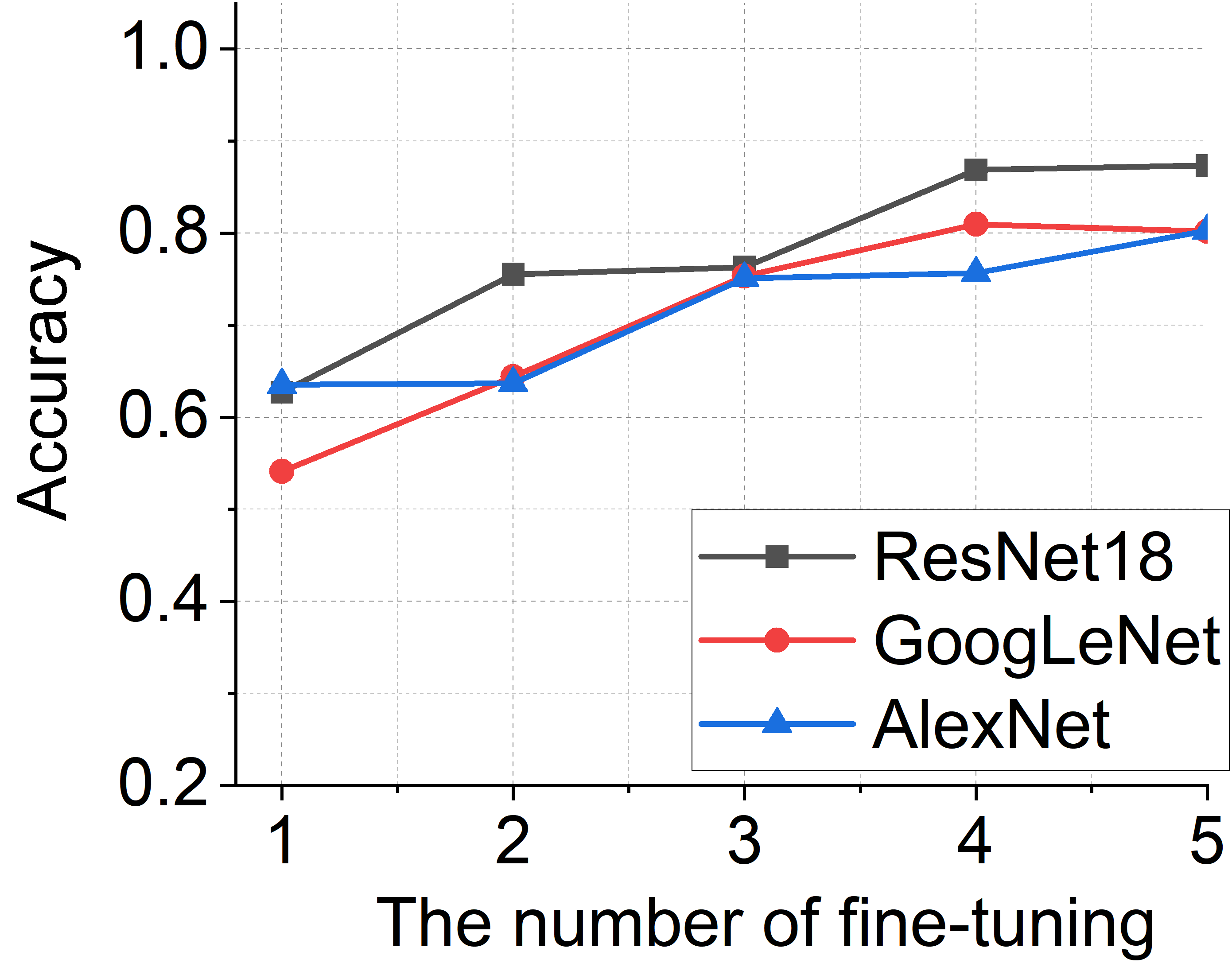}}
					\subfigure[CIFAR100]{\includegraphics[width=1.05in, height=0.9in]{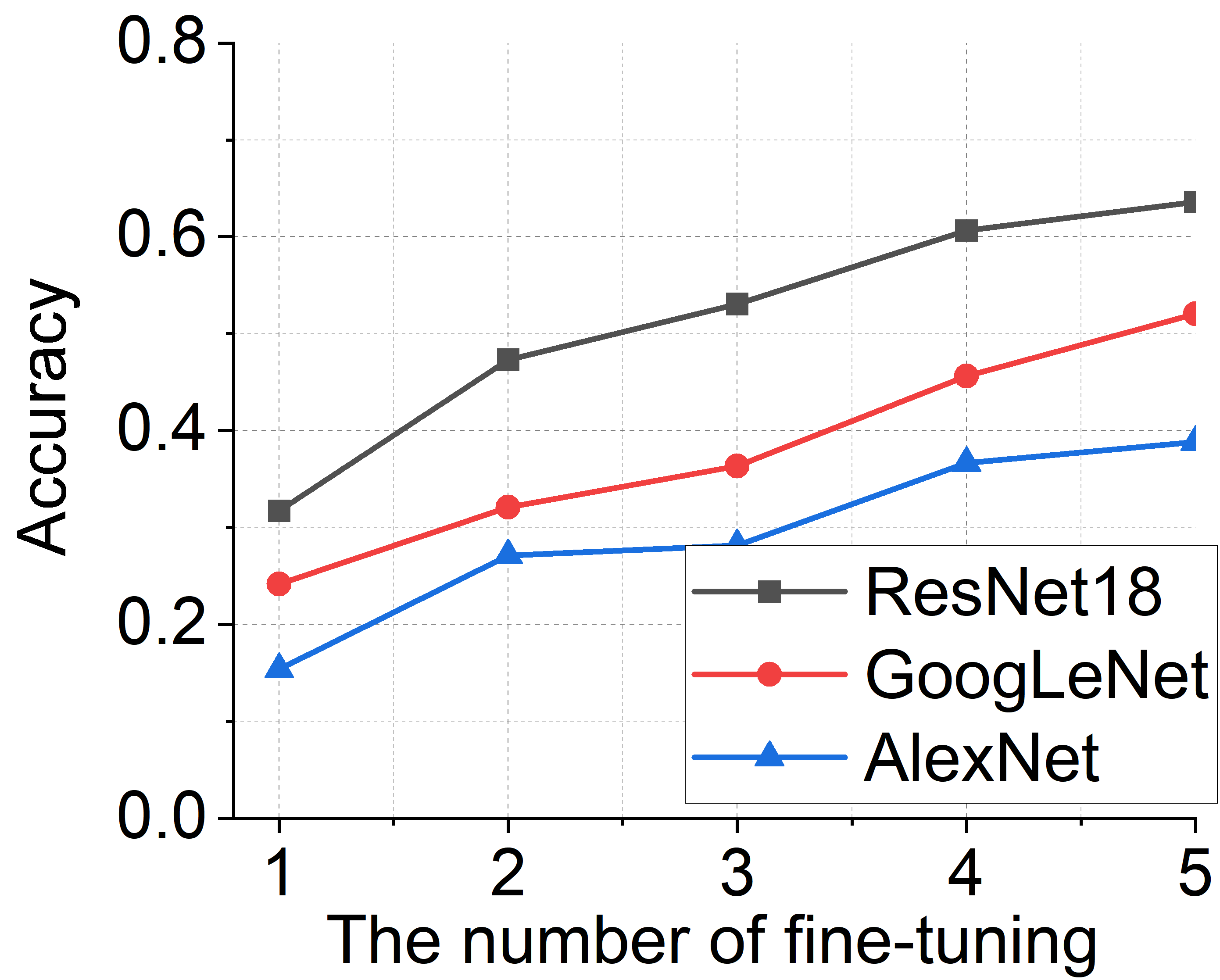}}
	\caption{The results of removal attacks.}
	\label{fig:finetuning}
\end{figure}

\begin{figure}[t]
	\centering
	\includegraphics[height=1.2in,width=1.7in]{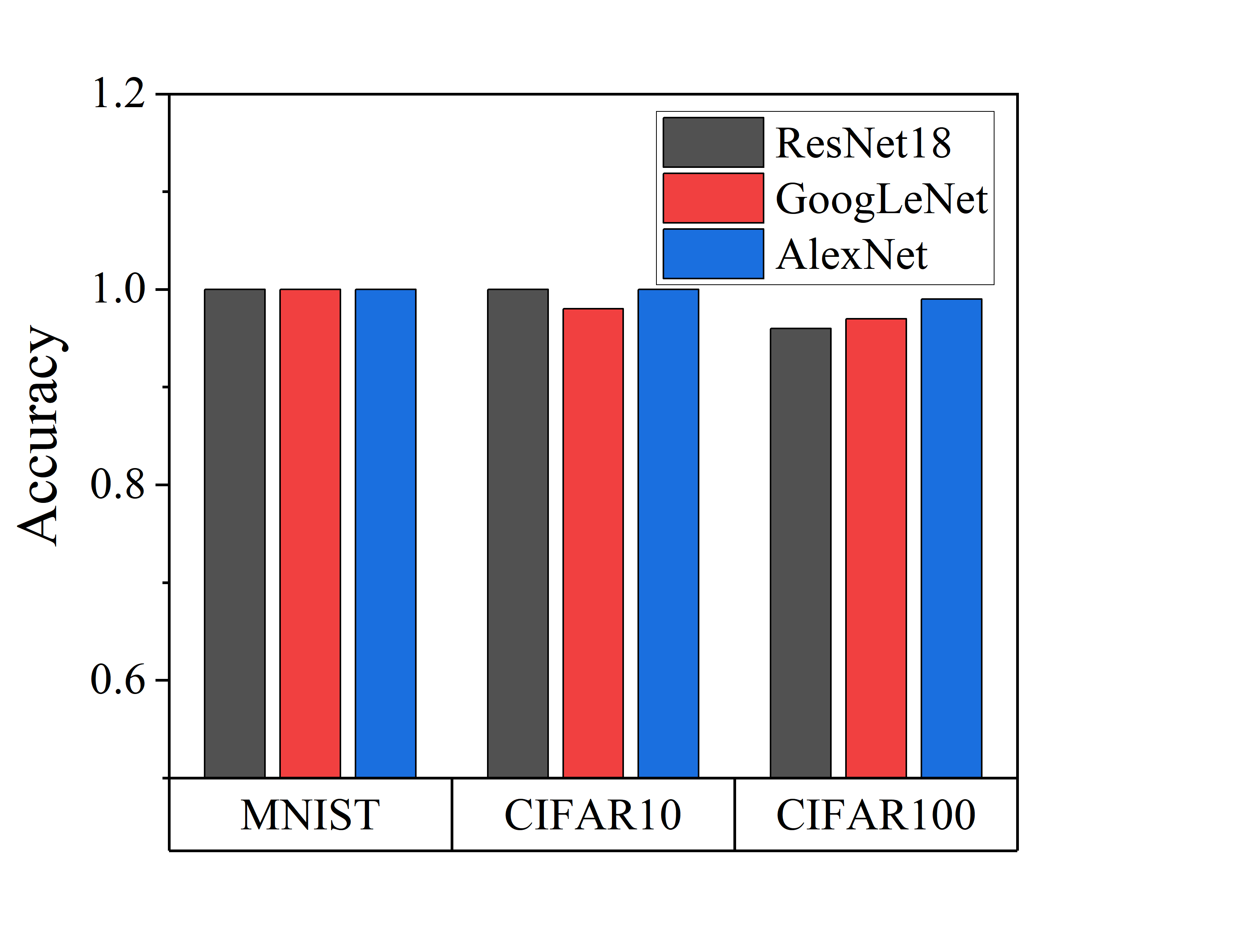}
	\caption{The verification on removal attacks.}
	\label{fig:vf_ft}
\end{figure} 

\subsection{Attack analysis} 
To evaluate removal attacks, we simulate an experiment involving the influence of fine-tuning. We partition each dataset into five parts and consecutively fine-tune the model on each part. We use the pre-trained model from the previous part as a starting point for the current training. The training procedure follows the Algorithm 1. Figure \ref{fig:finetuning} illustrates the trend in model performance after each fine-tuning round. The outcomes across the three datasets reveal that fine-tuning does not impact \textit{EncryIP}, and the model performance trend remains consistent within the expected range. To examine the verification, we conduct tests on its performance after each round of fine-tuning, employing the same experimental setup as described in Sec. \ref{vtfd}. Figure \ref{fig:vf_ft} presents the verification performance, which remains consistently close to 100\% accuracy across different model structures and datasets. These results demonstrate the robustness of  \textit{EncryIP} against this type of attack as well. 

\section{Conclusion}

This paper presents an innovative encryption-based framework for protecting model intellectual property. The framework employs label space encryption, establishing autonomy between learning and encryption algorithms. \textit{EncryIP} effectively ensures IP protection while maintaining efficiency across various model structures. Experimental results validate its effectiveness. Future work will focus on enhancing the framework's efficiency and security, exploring its theoretical underpinnings, and extending the concept to address dynamic IP protection challenges.

\section{Acknowledgments}

This work was supported by the National Key Research and Development Program of China (No. 2022ZD0115301), the National Natural Science Foundation of China (62106114, 62206139), the Major Key Project of PCL (PCL2023A09), National Natural Science Foundation of China under Grant No. U2001205, Guangdong Basic and Applied Basic Research Foundation (Grant No. 2023B1515040020),

\bibliography{aaai24}

\begin{thebibliography}{36}
\providecommand{\natexlab}[1]{#1}

\bibitem[{Adi et~al.(2018)Adi, Baum, Ciss{\'{e}}, Pinkas, and
  Keshet}]{DBLP:conf/uss/AdiBCPK18}
Adi, Y.; Baum, C.; Ciss{\'{e}}, M.; Pinkas, B.; and Keshet, J. 2018.
\newblock Turning Your Weakness Into a Strength: Watermarking Deep Neural
  Networks by Backdooring.
\newblock In \emph{{USENIX} Security Symposium}, 1615--1631.

\bibitem[{Boenisch(2020)}]{DBLP:journals/corr/abs-2009-12153}
Boenisch, F. 2020.
\newblock A Survey on Model Watermarking Neural Networks.
\newblock \emph{CoRR}, abs/2009.12153.

\bibitem[{Cao, Jia, and Gong(2021)}]{DBLP:conf/asiaccs/CaoJG21}
Cao, X.; Jia, J.; and Gong, N.~Z. 2021.
\newblock IPGuard: Protecting Intellectual Property of Deep Neural Networks via
  Fingerprinting the Classification Boundary.
\newblock In \emph{{ASIA} {CCS}}, 14--25.

\bibitem[{Chen et~al.(2018)Chen, Rouhani, Fan, Kilinc, and
  Koushanfar}]{DBLP:journals/corr/abs-1811-03713}
Chen, H.; Rouhani, B.~D.; Fan, X.; Kilinc, O.~C.; and Koushanfar, F. 2018.
\newblock Performance Comparison of Contemporary {DNN} Watermarking Techniques.
\newblock \emph{CoRR}, abs/1811.03713.

\bibitem[{Chen et~al.(2019)Chen, Rouhani, Fu, Zhao, and
  Koushanfar}]{DBLP:conf/mir/ChenRFZK19}
Chen, H.; Rouhani, B.~D.; Fu, C.; Zhao, J.; and Koushanfar, F. 2019.
\newblock DeepMarks: {A} Secure Fingerprinting Framework for Digital Rights
  Management of Deep Learning Models.
\newblock In \emph{ICMR}, 105--113.

\bibitem[{Chen, Rouhani, and
  Koushanfar(2019)}]{DBLP:journals/corr/abs-1904-00344}
Chen, H.; Rouhani, B.~D.; and Koushanfar, F. 2019.
\newblock BlackMarks: Blackbox Multibit Watermarking for Deep Neural Networks.
\newblock \emph{CoRR}, abs/1904.00344.

\bibitem[{Chen et~al.(2017)Chen, Liu, Li, Lu, and
  Song}]{DBLP:journals/corr/abs-1712-05526}
Chen, X.; Liu, C.; Li, B.; Lu, K.; and Song, D. 2017.
\newblock Targeted Backdoor Attacks on Deep Learning Systems Using Data
  Poisoning.
\newblock \emph{CoRR}, abs/1712.05526.

\bibitem[{Cramer and Shoup(1998)}]{CS98}
Cramer, R.; and Shoup, V. 1998.
\newblock A practical public key cryptosystem provably secure against adaptive
  chosen ciphertext attack.
\newblock In \emph{CRYPTO}, 13--25.

\bibitem[{Cramer and Shoup(2002)}]{CS02}
Cramer, R.; and Shoup, V. 2002.
\newblock Universal Hash Proofs and a Paradigm for Adaptive Chosen Ciphertext
  Secure Public-Key Encryption.
\newblock In \emph{EUROCRYPT}, 45--64.

\bibitem[{Fan, Ng, and Chan(2019)}]{DBLP:conf/nips/FanNC19}
Fan, L.; Ng, K.~W.; and Chan, C.~S. 2019.
\newblock Rethinking Deep Neural Network Ownership Verification: Embedding
  Passports to Defeat Ambiguity Attacks.
\newblock In \emph{NeurIPS}, 4716--4725.

\bibitem[{Fan et~al.(2022)Fan, Ng, Chan, and Yang}]{9454280}
Fan, L.; Ng, K.~W.; Chan, C.~S.; and Yang, Q. 2022.
\newblock DeepIPR: Deep Neural Network Ownership Verification With Passports.
\newblock \emph{IEEE TPAMI}, 44(10): 6122--6139.

\bibitem[{Feng and Zhang(2020)}]{DBLP:conf/ksem/FengZ20}
Feng, L.; and Zhang, X. 2020.
\newblock Watermarking Neural Network with Compensation Mechanism.
\newblock In \emph{KSEM}, 363--375.

\bibitem[{Guo and Potkonjak(2018)}]{DBLP:conf/iccad/GuoP18}
Guo, J.; and Potkonjak, M. 2018.
\newblock Watermarking deep neural networks for embedded systems.
\newblock In \emph{ICCAD}, 133--139.

\bibitem[{He et~al.(2016)He, Zhang, Ren, and Sun}]{DBLP:conf/cvpr/HeZRS16}
He, K.; Zhang, X.; Ren, S.; and Sun, J. 2016.
\newblock Deep Residual Learning for Image Recognition.
\newblock In \emph{CVPR}, 770--778.

\bibitem[{Hofheinz and Kiltz(2007)}]{HK07}
Hofheinz, D.; and Kiltz, E. 2007.
\newblock Secure Hybrid Encryption from Weakened Key Encapsulation.
\newblock In \emph{CRYPTO}, 553--571.

\bibitem[{Jebreel et~al.(2021)Jebreel, Domingo-Ferrer, Sánchez, and
  Blanco-Justicia}]{app11030999}
Jebreel, N.~M.; Domingo-Ferrer, J.; Sánchez, D.; and Blanco-Justicia, A. 2021.
\newblock KeyNet: An Asymmetric Key-Style Framework for Watermarking Deep
  Learning Models.
\newblock \emph{Applied Sciences}, 11(3).

\bibitem[{Katz and Lindell(2020)}]{KL20}
Katz, J.; and Lindell, Y. 2020.
\newblock \emph{Introduction to modern cryptography}.
\newblock CRC press.

\bibitem[{Krizhevsky, Sutskever, and
  Hinton(2012)}]{DBLP:conf/nips/KrizhevskySH12}
Krizhevsky, A.; Sutskever, I.; and Hinton, G.~E. 2012.
\newblock ImageNet Classification with Deep Convolutional Neural Networks.
\newblock In \emph{NIPS}, 1106--1114.

\bibitem[{Kuribayashi, Tanaka, and
  Funabiki(2020)}]{DBLP:conf/apsipa/KuribayashiTF20}
Kuribayashi, M.; Tanaka, T.; and Funabiki, N. 2020.
\newblock DeepWatermark: Embedding Watermark into {DNN} Model.
\newblock In \emph{APSIPA}, 1340--1346.

\bibitem[{Lukas, Zhang, and Kerschbaum(2021)}]{DBLP:conf/iclr/LukasZK21}
Lukas, N.; Zhang, Y.; and Kerschbaum, F. 2021.
\newblock Deep Neural Network Fingerprinting by Conferrable Adversarial
  Examples.
\newblock In \emph{ICLR}.

\bibitem[{Maung and Kiya(2020)}]{DBLP:conf/gcce/PyoneMK20}
Maung, A. P.~M.; and Kiya, H. 2020.
\newblock Training {DNN} Model with Secret Key for Model Protection.
\newblock In \emph{{IEEE} {GCCE}}, 818--821.

\bibitem[{Merrer, P{\'{e}}rez, and
  Tr{\'{e}}dan(2020)}]{DBLP:journals/nca/MerrerPT20}
Merrer, E.~L.; P{\'{e}}rez, P.; and Tr{\'{e}}dan, G. 2020.
\newblock Adversarial frontier stitching for remote neural network
  watermarking.
\newblock \emph{Neural Computing \& Applications}, 32(13): 9233--9244.

\bibitem[{Naor and Segev(2009)}]{NS09}
Naor, M.; and Segev, G. 2009.
\newblock Public-Key Cryptosystems Resilient to Key Leakage.
\newblock In \emph{CRYPTO}, 18--35.

\bibitem[{OpenAI(2023)}]{DBLP:journals/corr/abs-2303-08774}
OpenAI. 2023.
\newblock {GPT-4} Technical Report.
\newblock \emph{CoRR}, abs/2303.08774.

\bibitem[{Paillier(1999)}]{Paillier99}
Paillier, P. 1999.
\newblock Public-Key Cryptosystems Based on Composite Degree Residuosity
  Classes.
\newblock In \emph{EUROCRYPT}, 223--238.
\newblock ISBN 978-3-540-48910-8.

\bibitem[{Ramesh et~al.(2021)Ramesh, Pavlov, Goh, Gray, Voss, Radford, Chen,
  and Sutskever}]{DBLP:conf/icml/RameshPGGVRCS21}
Ramesh, A.; Pavlov, M.; Goh, G.; Gray, S.; Voss, C.; Radford, A.; Chen, M.; and
  Sutskever, I. 2021.
\newblock Zero-Shot Text-to-Image Generation.
\newblock In \emph{{ICML}}.

\bibitem[{Rouhani, Chen, and Koushanfar(2019)}]{DBLP:conf/asplos/RouhaniCK19}
Rouhani, B.~D.; Chen, H.; and Koushanfar, F. 2019.
\newblock DeepSigns: An End-to-End Watermarking Framework for Ownership
  Protection of Deep Neural Networks.
\newblock In \emph{ASPLOS}, 485--497.

\bibitem[{Szegedy et~al.(2015)Szegedy, Liu, Jia, Sermanet, Reed, Anguelov,
  Erhan, Vanhoucke, and Rabinovich}]{DBLP:conf/cvpr/SzegedyLJSRAEVR15}
Szegedy, C.; Liu, W.; Jia, Y.; Sermanet, P.; Reed, S.~E.; Anguelov, D.; Erhan,
  D.; Vanhoucke, V.; and Rabinovich, A. 2015.
\newblock Going deeper with convolutions.
\newblock In \emph{CVPR}, 1--9.

\bibitem[{Tauhid et~al.(2023)Tauhid, Xu, Rahman, and Tomai}]{TAUHID2023100114}
Tauhid, A.; Xu, L.; Rahman, M.; and Tomai, E. 2023.
\newblock A survey on security analysis of machine learning-oriented hardware
  and software intellectual property.
\newblock \emph{High-Confidence Computing}, 3(2): 100114.

\bibitem[{Uchida et~al.(2017)Uchida, Nagai, Sakazawa, and
  Satoh}]{DBLP:conf/mir/UchidaNSS17}
Uchida, Y.; Nagai, Y.; Sakazawa, S.; and Satoh, S. 2017.
\newblock Embedding Watermarks into Deep Neural Networks.
\newblock In \emph{ICMR}, 269--277.

\bibitem[{Xue, Wang, and Liu(2021)}]{DBLP:conf/glvlsi/Xue0L21}
Xue, M.; Wang, J.; and Liu, W. 2021.
\newblock {DNN} Intellectual Property Protection: Taxonomy, Attacks and
  Evaluations.
\newblock In \emph{Great Lakes Symposium on {VLSI}}, 455--460.

\bibitem[{Zhang et~al.(2020{\natexlab{a}})Zhang, Chen, Liao, Fang, Zhang, Zhou,
  Cui, and Yu}]{DBLP:conf/aaai/ZhangCLFZZCY20}
Zhang, J.; Chen, D.; Liao, J.; Fang, H.; Zhang, W.; Zhou, W.; Cui, H.; and Yu,
  N. 2020{\natexlab{a}}.
\newblock Model Watermarking for Image Processing Networks.
\newblock In \emph{AAAI}, 12805--12812.

\bibitem[{Zhang et~al.(2022)Zhang, Chen, Liao, Zhang, Feng, Hua, and
  Yu}]{DBLP:journals/pami/ZhangCLZFHY22}
Zhang, J.; Chen, D.; Liao, J.; Zhang, W.; Feng, H.; Hua, G.; and Yu, N. 2022.
\newblock Deep Model Intellectual Property Protection via Deep Watermarking.
\newblock \emph{IEEE TPAMI}, 44(8): 4005--4020.

\bibitem[{Zhang et~al.(2020{\natexlab{b}})Zhang, Chen, Liao, Zhang, Hua, and
  Yu}]{DBLP:conf/nips/Zhang00Z0Y20}
Zhang, J.; Chen, D.; Liao, J.; Zhang, W.; Hua, G.; and Yu, N.
  2020{\natexlab{b}}.
\newblock Passport-aware Normalization for Deep Model Protection.
\newblock In \emph{NeurIPS}.

\bibitem[{Zhang et~al.(2018)Zhang, Gu, Jang, Wu, Stoecklin, Huang, and
  Molloy}]{DBLP:conf/ccs/ZhangGJWSHM18}
Zhang, J.; Gu, Z.; Jang, J.; Wu, H.; Stoecklin, M.~P.; Huang, H.; and Molloy,
  I.~M. 2018.
\newblock Protecting Intellectual Property of Deep Neural Networks with
  Watermarking.
\newblock In \emph{ASIA CCS}, 159--172.

\bibitem[{Zhong et~al.(2020)Zhong, Zhang, Zhang, Gao, and
  Xiang}]{DBLP:conf/pakdd/ZhongZ0G020}
Zhong, Q.; Zhang, L.~Y.; Zhang, J.; Gao, L.; and Xiang, Y. 2020.
\newblock Protecting {IP} of Deep Neural Networks with Watermarking: {A} New
  Label Helps.
\newblock In \emph{{PAKDD}}, 462--474.

\end{thebibliography}


\appendix
\section{More Definitions}\label{sec:appendix_definition}
In this paper, we employ a simplified version of the Cramer-Shoup  encryption scheme \cite{CS98} as a building block, where the simplified scheme achieves \emph{ciphertext indistinguishability under chosen-plaintext attacks  (IND-CPA security)} under the well-studied decisional Diffie-Hellman (DDH) assumption. For completeness, we recall the definition of the DDH assumption and that of IND-CPA security as follows. 
\begin{definition}{(The DDH assumption).}
	Let $\mathbb{G}_q$ be a cyclic group of prime order $q$. We say that \emph{the decisional Diffie-Hellman (DDH) assumption holds for $\mathbb{G}_q$}, if for any probabilistic polynomial-time  distinguisher $\D$, its advantage 
	\begin{align*}
		\mathbf{Adv}_{\mathbb{G}_q,\D}^{\textup{ddh}}=&|\Pr{[\D(g,g^a,g^b,g^{ab})=1]}\\
		&~~~~~~~~~~~~~~-\Pr{[\D(g,g^a,g^b,g^{c})=1]}|
	\end{align*}
	is negligible\footnote{For the definition of \emph{negligible probability}, please refer to \cite{KL20}.}, where $g$ is uniformly sampled from $\mathbb{G}_q\setminus\{1\}$, and $a,b,c$ are all uniformly sampled from $\mathbb{Z}_q$.
	\label{DDH}
\end{definition}
\begin{definition}{(IND-CPA security).}
	We say that a PKE scheme $\pke=(\gen,\enc,\dec,\fake)$ is  \emph{IND-CPA secure}, if for any probabilistic polynomial-time adversary $\A=(\A_1,\A_2)$, its advantage 
	\begin{align*}
		\mathbf{Adv}_{\pke,\A}^{\textup{ind-cpa}}=|\Pr{[\mathbf{Exp}_{\pke,\A}^{\textup{ind-cpa}}=1]}-\frac{1}{2}|
	\end{align*}
	is negligible, where game $\mathbf{Exp}_{\pke,\A}^{\textup{ind-cpa}}$ is defined in Figure \ref{fig:IND-CPA_game}.
	\label{def:IND-CPA}
	\begin{figure}[!htb]  
		\small
		\centering
		\begin{tabular}[c]{|l|}\hline
			\underline{$\mathbf{Exp}_{\pke,\A}^{\textup{ind-cpa}}$:}  \\
			
			$(pk,\{sk_j\}_{j=1}^{P})\leftarrow\textsf{\upshape{Gen}}(P)$, $\beta\leftdollar\{0,1\}$  \\
			
			$(m_0,m_1,st_{\A})\leftarrow\A_1(pk)$ (s.t. $|m_0|=|m_1|$) 	   \\

			$c\leftarrow\enc(pk,m_{\beta})$, $\beta'\leftarrow\A_2(c,st_{\A})$	 \\
			
			If $\beta'=\beta$: Return $1$; $~$ Else Return $0$\\\hline
		\end{tabular}
		\caption{Game defining IND-CPA security for PKE.}\label{fig:IND-CPA_game}
	\end{figure}
\end{definition}

Note that IND-CPA security is much stronger than one-wayness. It requires that, informally, after the encryption, a ciphertext will not leak even a single-bit information about the encrypted message,  except for the message length.

\section{Correctness Analysis of $\csl$}\label{sec:appendix_correctness_analysis}
Correctness analysis of the PKE  scheme $\csl$ is as follows. 
\begin{enumerate}
	\item[(i)] For any ciphertext  $c=(u_1,u_2,u_3)\leftarrow\enc(pk,m)$, decrypting it with secret key $sk_1=(a_1,b_1)$, we will obtain
		\[\dec(sk_1,c)=\frac{u_3}{u_1^{a_1}u_2^{b_1}}=\frac{h^rm}{g_1^{ra_1}g_2^{rb_1}}=\frac{(g_1^{a_1}g_2^{b_1})^rm}{g_1^{ra_1}g_2^{rb_1}}=m.\]
	Note that for $2\leq j\leq P$, 
		\[g_1^{a_j}g_2^{b_j}=g_1^{a_1+(b_1-b_j)t}g_2^{b_j}=g_1^{a_1}g_2^{(b_1-b_j)}g_2^{b_j}=g_1^{a_1}g_2^{b_1}=h.\]
	So decrypting the ciphertext with $sk_j$, we will obtain
		\[\dec(sk_j,c)=\frac{u_3}{u_1^{a_j}u_2^{b_j}}=\frac{h^rm}{g_1^{ra_j}g_2^{rb_j}}=\frac{(g_1^{a_j}g_2^{b_j})^rm}{g_1^{ra_j}g_2^{rb_j}}=m.\]
	\item[(ii)] For any ill-formed ciphertext $c=(u_1,u_2,u_3)\leftarrow\fake(pk)$, decrypting it with secret key $sk_j=(a_j,b_j)$ where $j\neq 1$, since 
	\[g_1^{r_1a_j}g_2^{r_2b_j}=g_1^{r_1(a_1+(b_1-b_j)t)}g_2^{r_2b_j}=g_1^{r_1a_1}g_2^{r_1b_1}g_2^{(r_2-r_1)b_j},\] 
	we will obtain
		\[\dec(sk_j,c)=\frac{u_3}{g_1^{r_1a_j}g_2^{r_2b_j}}=\frac{u_3}{g_1^{r_1a_1}g_2^{r_1b_1}}\cdot\frac{1}{g_2^{(r_2-r_1)b_j}}.\]
	Note that $r_1\neq r_2$, and $b_{j_1}\neq b_{j_2}$ for any $j_1\neq j_2$, so we conclude that $\dec(sk_{j_1},c)\neq \dec(sk_{j_2},c)$. 
\end{enumerate}

\section{Proof of Theorem \ref{thm:CPA}}\label{sec:appendix_proof}
This proof is similar to that in \cite{CS98}. 
For any probabilistic polynomial-time adversary $\A=(\A_1,\A_2)$ attacking the IND-CPA security of $\csl$, we show a probabilistic polynomial-time distinguisher $\D$ attacking  the DDH assumption as in Figure \ref{fig:distinguisher}.
\begin{figure}[!htb]  
	\small
	\centering
	\begin{tabular}[c]{|l|}\hline
		\underline{$\D(g,g^a,g^b,g^z)$:}  \\
		
		$g_1=g$, $g_2=g^a$, $(a_1,b_1)\leftdollar(\mathbb{Z}_q)^2$, $h=g_1^{a_1}g_2^{b_1}$\\
		
		$pk=(g_1,g_2,h)$, $(m_0,m_1,st_{\A})\leftarrow\A_1(pk)$ (s.t. $|m_0|=|m_1|$)  \\
		
		$\beta\leftdollar\{0,1\}$,  $u_1=g^b$, $u_2=g^z$, $u_3=u_1^{a_1}u_2^{b_1}m_{\beta}$ 	   \\
		
		$c=(u_1,u_2,u_3)$, 		
		$\beta'\leftarrow\A_2(c,st_{\A})$	 \\
		
		If $\beta'=\beta$: Return $1$; $~$ Else: Return $0$\\\hline
	\end{tabular}
	\caption{Distinguisher $\D$ in the proof of Theorem \ref{thm:CPA}.}\label{fig:distinguisher}
\end{figure}

We analyze $\D$'s advantage as follows. 

When $g^{z}=g^{ab}$, we derive that (i) the constructed public key has the form $pk=(g_1,g_2,h=g_1^{a_1}g_2^{b_1})$, and (ii) for the constructed ciphertext $c=(u_1,u_2,u_3)$, $u_1=g^b=g_1^b$, $u_2=g^z=g^{ab}=g_2^{b}$, and $u_3=u_1^{a_1}u_2^{b_1}m_{\beta}=(g_1^b)^{a_1}(g_2^{b})^{b_1}m_{\beta}=(g_1^{a_1}g_2^{b_1})^{b}m_{\beta}=h^bm_{\beta}$. So in this case, the distribution of the public key and the ciphertext correspond exactly to $\A$'s view in game $\mathbf{Exp}_{\pke,\A}^{\textup{ind-cpa}}$. Hence, 
\begin{small}
	\begin{eqnarray}\label{eq:pf_1}
		\Pr[\D(g,g^a,g^b,g^{ab})=1]=\Pr[\mathbf{Exp}_{\pke,\A}^{\textup{ind-cpa}}=1].
	\end{eqnarray}
\end{small}

When $g^{z}=g^{c}$, we have $c\neq ab$ with all but negligible probability (i.e., $\frac{1}{q}$), since $c$ is uniformly and independently sampled from $\mathbb{Z}_q$. From now on, we simply assume  $c\neq ab$.

In this case, denote by $\mu=u_1^{a_1}u_2^{b_1}=g^{ba_1+cb_1}$. 
Given the public key $pk$ and the ciphertext $c$, from $\A$'s point of view, what information that  $\A$ has about $(a_1,b_1)$ is 
\begin{small}
	\begin{eqnarray*}
		\log_{g}h=a_1+ab_1\ \ \ \ \ \text{and}\ \ \ \ \ \log_g\mu=ba_1+cb_1.
	\end{eqnarray*}
\end{small}
These two equations 
are linearly independent since $c\neq ab$. So they have a unique solution in $a_1,b_1$. In other words, assuming that $c\neq ab$, given $pk=(g_1,g_2,h)$, from $\A$'s point of view, each possible value of $\mu$ leads to a possible value of $(a_1,b_1)$.  Note that $a_1$ and $b_1$ are both uniformly and independently sampled from $\mathbb{Z}_q$, so  from $\A$'s point of view, $\mu=u_1^{a_1}u_2^{b_1}$ is uniformly distributed, i.e., $\A$ has no information about which message was encrypted. Thus, the probability that $\A$ outputs $\beta'$ satisfying $\beta'=\beta$ is $\frac{1}{2}$. Hence, 
\begin{small}
	\begin{eqnarray}\label{eq:pf_4}
		\Pr[\D(g,g^a,g^b,g^{c})=1]=\frac{1}{2}\pm\textsf{negl},
	\end{eqnarray}
\end{small}
where $\textsf{negl}$ denotes a negligible value.

Combining equations (\ref{eq:pf_1}) and $(\ref{eq:pf_4})$, we obtain
\begin{small}
	\begin{eqnarray*}
		\mathbf{Adv}_{\mathbb{G}_q,\D}^{\textup{ddh}}&=&|\Pr{[\D(g,g^a,g^b,g^{ab})=1]}\\
		&&-\Pr{[\D(g,g^a,g^b,g^{c})=1]}|\\
		&=&\mathbf{Adv}_{\pke,\A}^{\textup{ind-cpa}}\pm\textsf{negl}.
	\end{eqnarray*}
\end{small}

So the DDH assumption guarantees that $\mathbf{Adv}_{\pke,\A}^{\textup{ind-cpa}}$ is negligible. \hfill$\square$

\section{Experiment Setup}
\label{ESp}


{\bf Data Sets and Models.} We use three datasets to compare the performance of all methods: {\bf MNIST}, {\bf CIFAR-10} and {\bf CIFAR-100}. The model architectures include ResNet18 \cite{DBLP:conf/cvpr/HeZRS16}, GoogLeNet \cite{DBLP:conf/cvpr/SzegedyLJSRAEVR15} and AlexNet \cite{DBLP:conf/nips/KrizhevskySH12}.  

\vspace{1mm}

\noindent {\bf Competing Algorithms.} The three typical methods from watermark-based (\textbf{PR} \cite{DBLP:conf/mir/UchidaNSS17}), trigger-based (\textbf{EW} \cite{DBLP:conf/ccs/ZhangGJWSHM18}) and passport-based method (\textbf{DeepIPR} \cite{9454280}) are chosen as the comparisons. 
\begin{itemize}[leftmargin=2em]
	
	\item \textbf{PR} \cite{DBLP:conf/mir/UchidaNSS17}: The method is a general framework for embedding a watermark in model parameters by using an embedding regularizer, e.g., $L(w) = (b_j \log(y_j) + (1 - b_j)\log(1 - y_j))$, $y_j=X_j w$, and  $X_j$ is an arbitrary matrix. 
	\item \textbf{EW} \cite{DBLP:conf/ccs/ZhangGJWSHM18}: The framework assigns pre-defined labels for different watermarks and trains the watermarks with pre-defined labels to DNNs. The DNNs automatically learn and memorize the patterns of embedded watermarks and pre-defined labels.
	
	\item \textbf{DeepIPR} \cite{9454280}: A passport-based verification scheme controls the DNN model functionalities by the embedded digital signatures i.e. passports. It introduces a new DNN layer (passport layer) by appending after a convolution layer.
\end{itemize}

	
	

\vspace{1mm}
\noindent{\bf Experiment Settings.} All experiments are implemented in Python on Intel Core CPU machine with 128 GB memory and NVIDIA RTX 2080 GPU. We use the public codes for both \textbf{PR} and \textbf{DeepIPR}, and the code of \textbf{EW} is developed based on the original papers. The common parameters in each model are set by default values. 
\vspace{1mm}

\noindent{\bf Evaluation Measures.} Without loss of generality, we employ the ``Accuracy (Acc.)'' as a basic metric to monitor how will these IP protection mechanisms affect original model performance.

\section{The convergence during training}
\label{cdt}
\begin{figure}[!htb]
	\centering
	\subfigure[MNIST]{\epsfig{file=resnet_mnist1.eps,height=0.9in,width=1.05in}} 
	\subfigure[CIFAR10]{\epsfig{file=resnet_cifar101.eps,height=0.9in,width=1.05in}} 
	\subfigure[CIFAR100]{\epsfig{file=resnet_cifar1001.eps,height=0.9in,width=1.05in}} 
	\subfigure[MNIST]{\epsfig{file=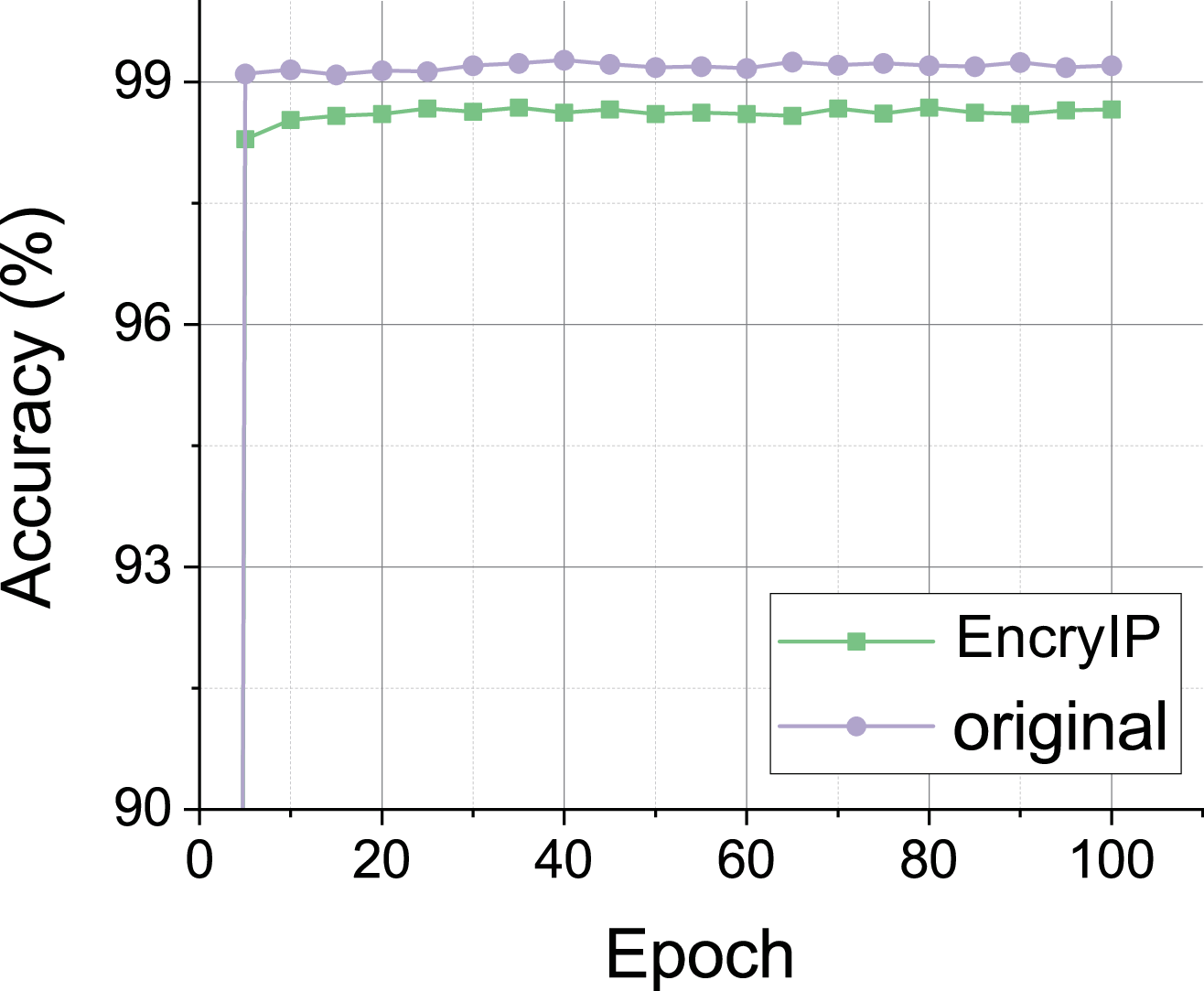,height=0.9in,width=1.05in}} 
	\subfigure[CIFAR10 ]{\epsfig{file=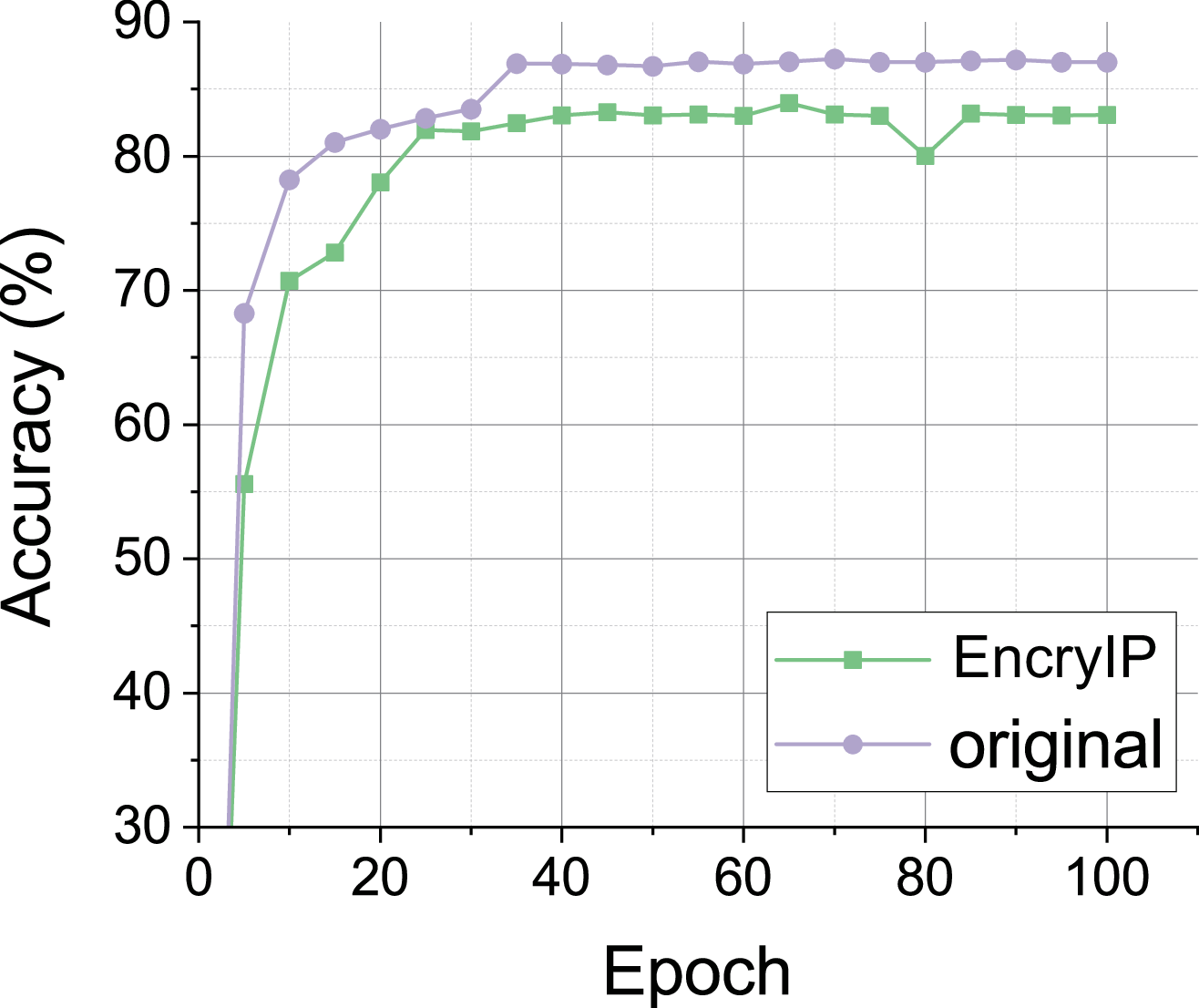,height=0.9in,width=1.05in}} 
	\subfigure[CIFAR100]{\epsfig{file=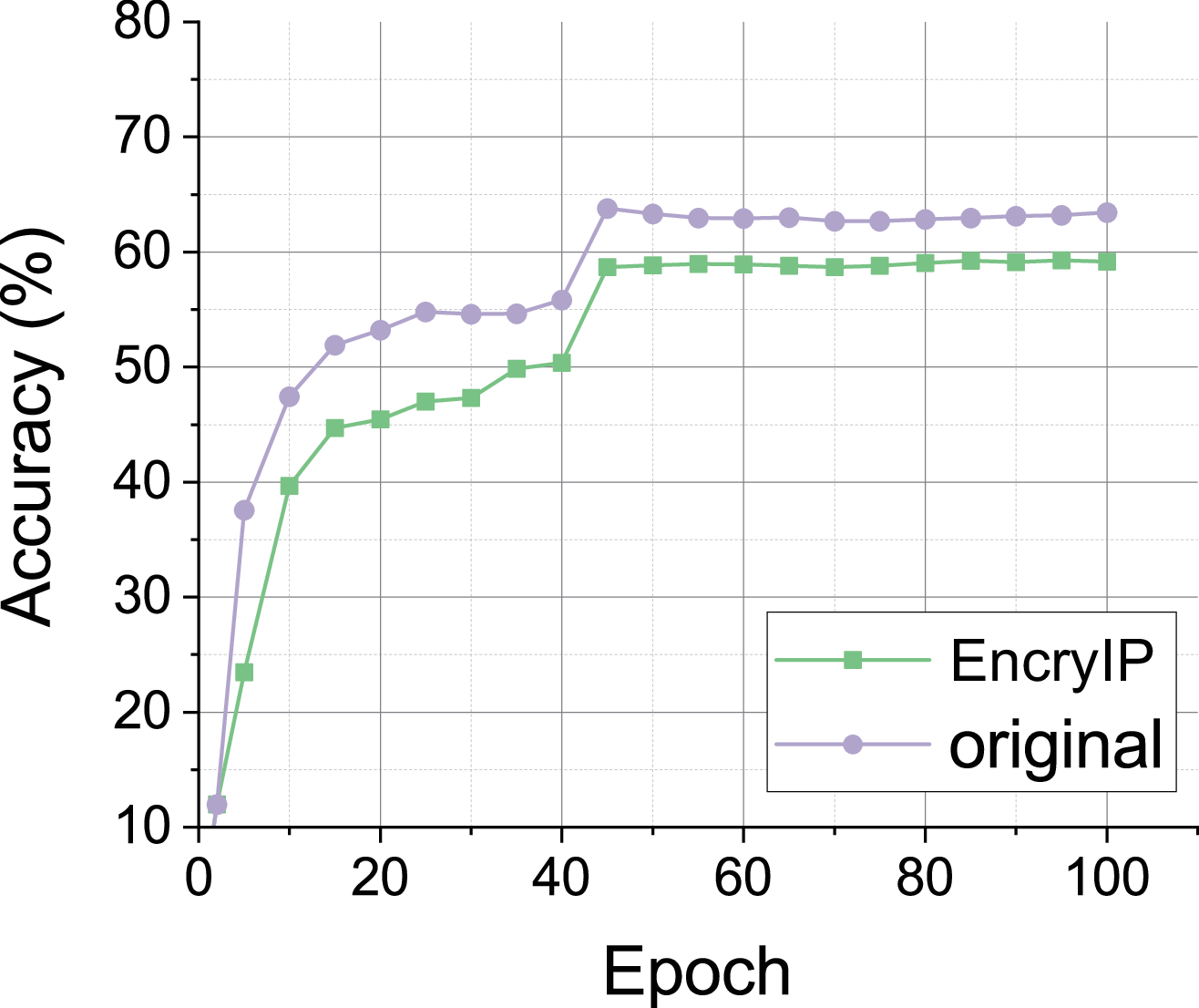,height=0.9in,width=1.05in}} 
	\subfigure[MNIST]{\epsfig{file=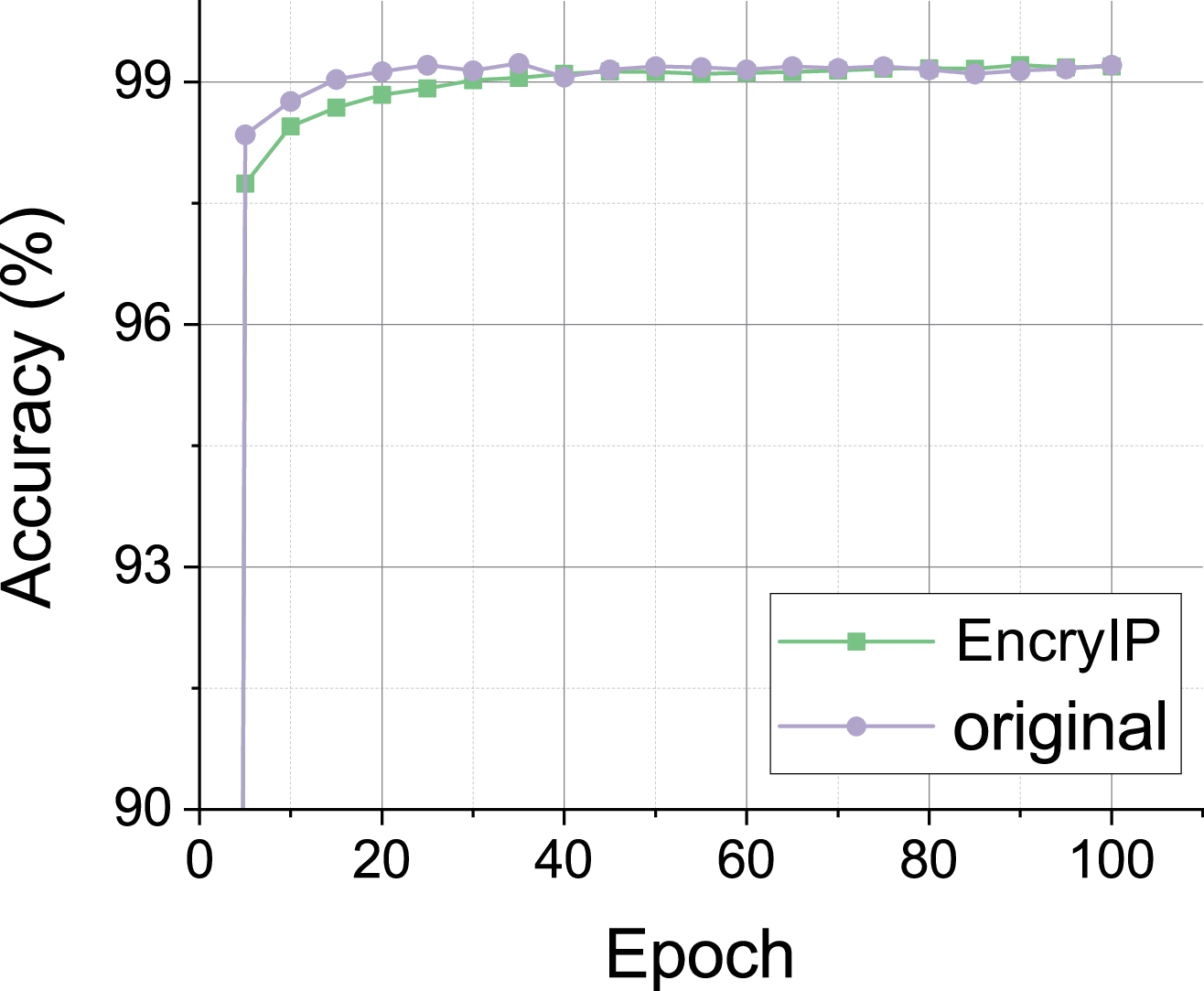,height=0.9in,width=1.05in}} 
	\subfigure[ACIFAR10]{\epsfig{file=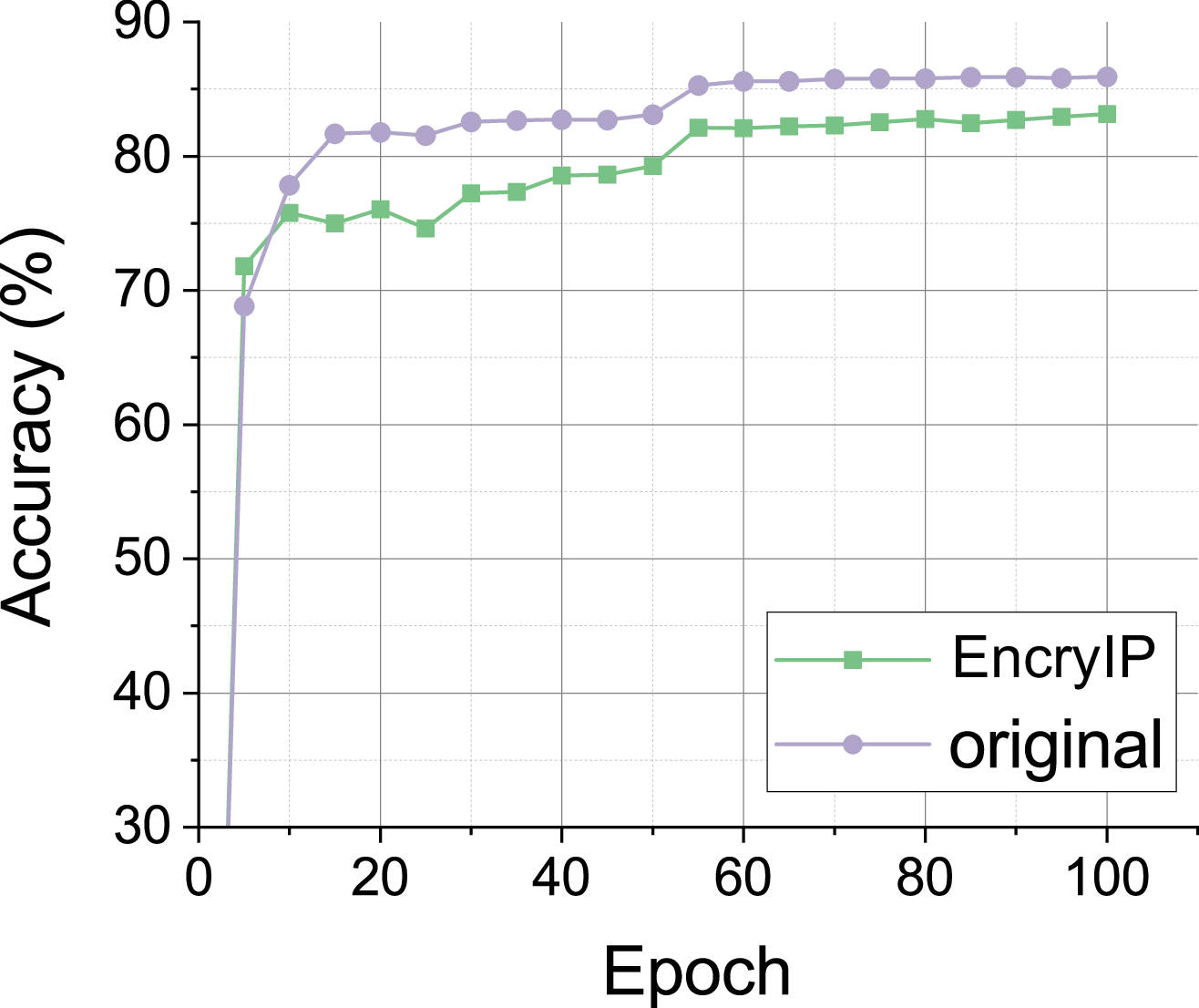,height=0.9in,width=1.05in}} 
	\subfigure[CIFAR100]{\epsfig{file=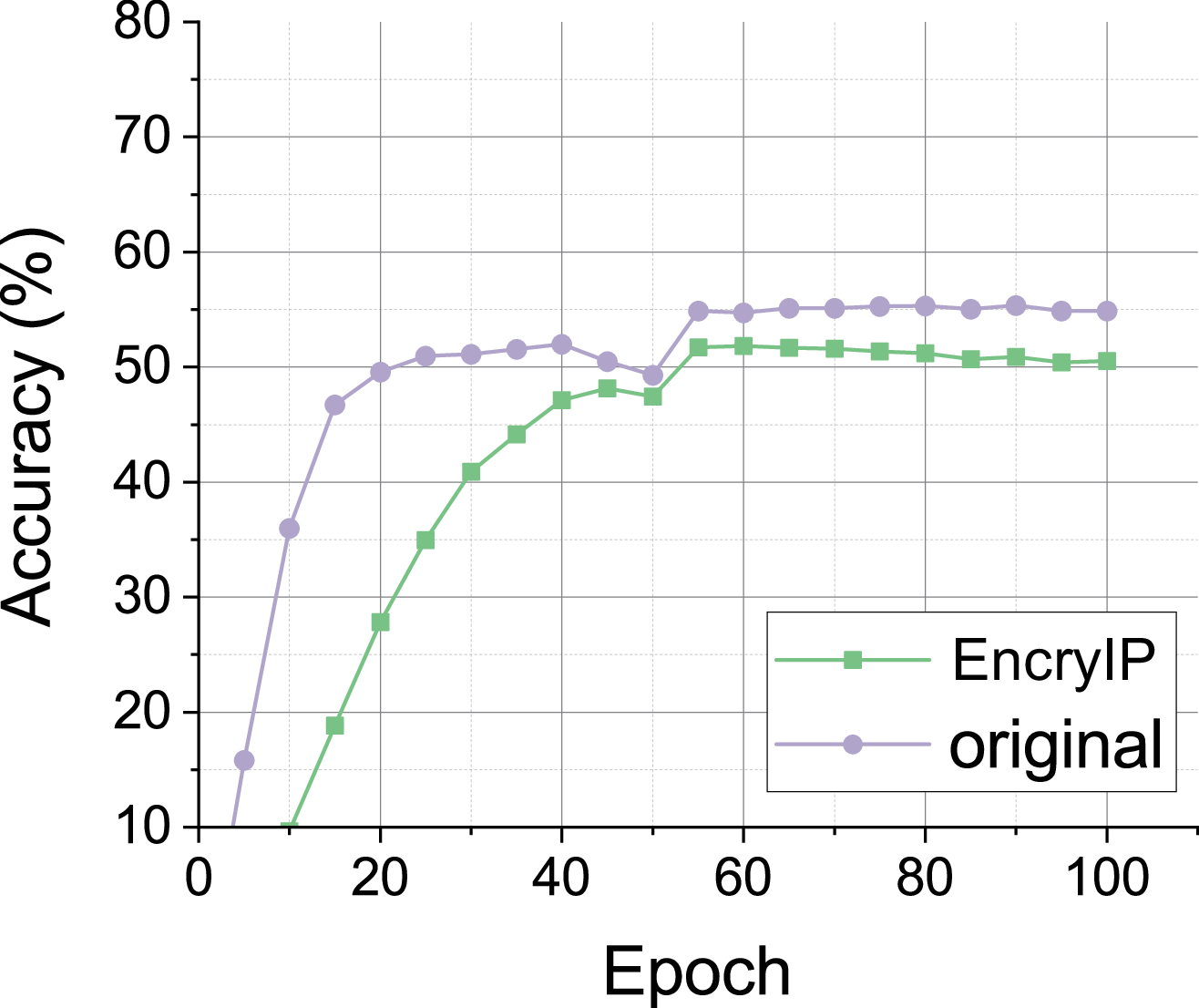,height=0.9in,width=1.05in}} 
	\caption{The convergence during training. (a)-(c): ResNet, (d)-(f): GoogLeNet, (g)-(i): AlexNet.}
	\label{fig:covergence1}
\end{figure}

The convergence during training is shown in Figure \ref{fig:covergence1}. 

\section{The efficiency of \textit{EncryIP}}
\label{fullresult}

The complete result of the efficiency is shown in Table \ref{tab:time}.

\begin{table}[!htb]
	\scriptsize
	\centering
	\caption{The results of training time (min.).}
	\begin{tabular}{ccccc}
		\toprule
		Model & Method &  MNIST & CIFAR10 &CIFAR100\\ 
		\midrule
		\multirow{5}*{ResNet18}& original & $17.61$ & $22.21$ & $56.27$ \\ 
		\multirow{5}*{} & PR  &53.4  &67.53 & 166.67\\ 
		\multirow{5}*{}& EW  &17.75  &24.02  & $66.31$\\ 
		\multirow{5}*{}& DeepIPR  & 55.03& $66.35$ & $170.26$ \\ 
		\multirow{5}*{} & \textit{EncryIP}  & \textbf{17.62} & \textbf{23.36} & \textbf{66.02} \\ 
		\midrule
		\multirow{5}*{GoogLeNet}& original & $11.22$ & $41.8$ & $78.47$ \\ 
		& PR  & 35.1& 122.21&  239.18\\ 
		& EW  & $12.67$ &$42.67$  & $83.15$ \\ 
		& DeepIPR  &35.63  &124.25 & 247.33 \\ 
		& \textit{EncryIP}  & \textbf{12.02} & \textbf{42.05} & \textbf{83.12} \\
		\midrule
		\multirow{5}*{AlexNet}& original & $8.75$ & $18.85$ & $21.08$ \\ 
		& PR  & 25.02 &54.32 &59.70 \\ 
		& EW  & 9.50 & $23.39$ & $24.42$\\ 
		& DeepIPR  &23.47  & $60.93$ & $60.62$ \\ 
		& \textit{EncryIP}  & \textbf{9.18} & \textbf{23.33} & \ \textbf{24.36 } \\ 
		\bottomrule
	\end{tabular}
	\label{tab:time}
\end{table}

\section{Parameter analysis} 

\begin{figure}[t]
	\centering
	\subfigure[]{\epsfig{file=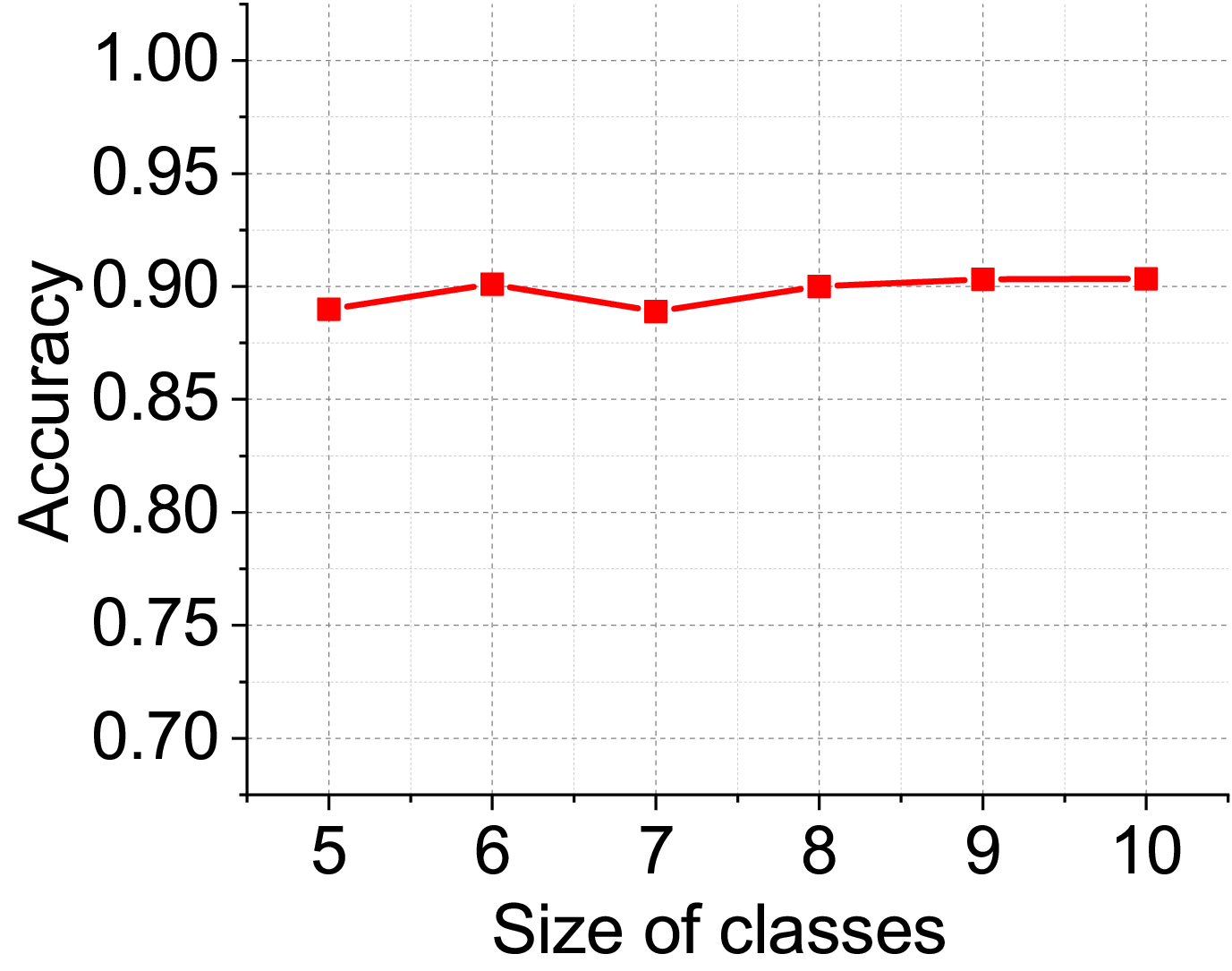,height=0.9in,width=1.05in}} 
	\subfigure[]{\epsfig{file=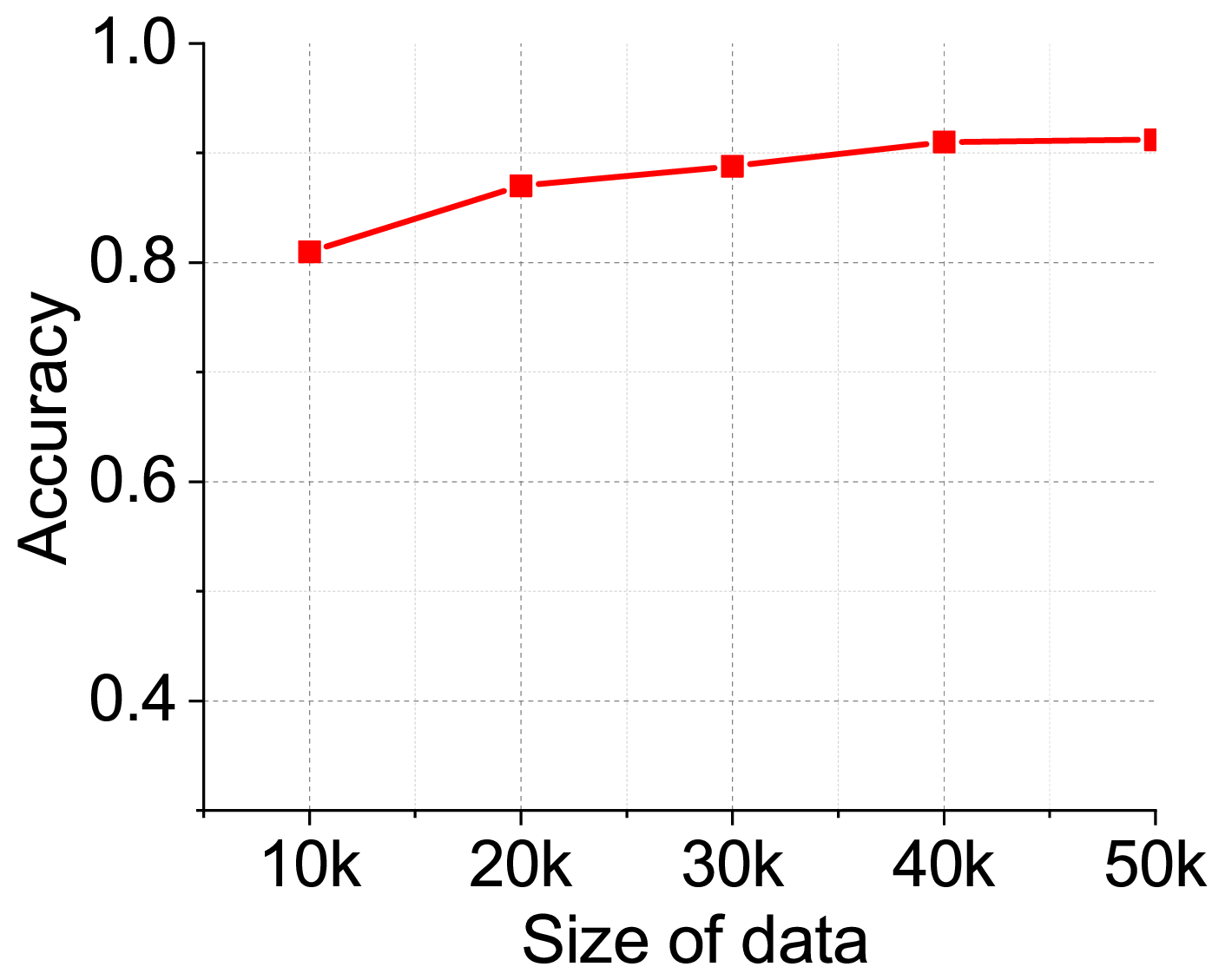,height=0.9in,width=1.05in}}
	\subfigure[]{\epsfig{file=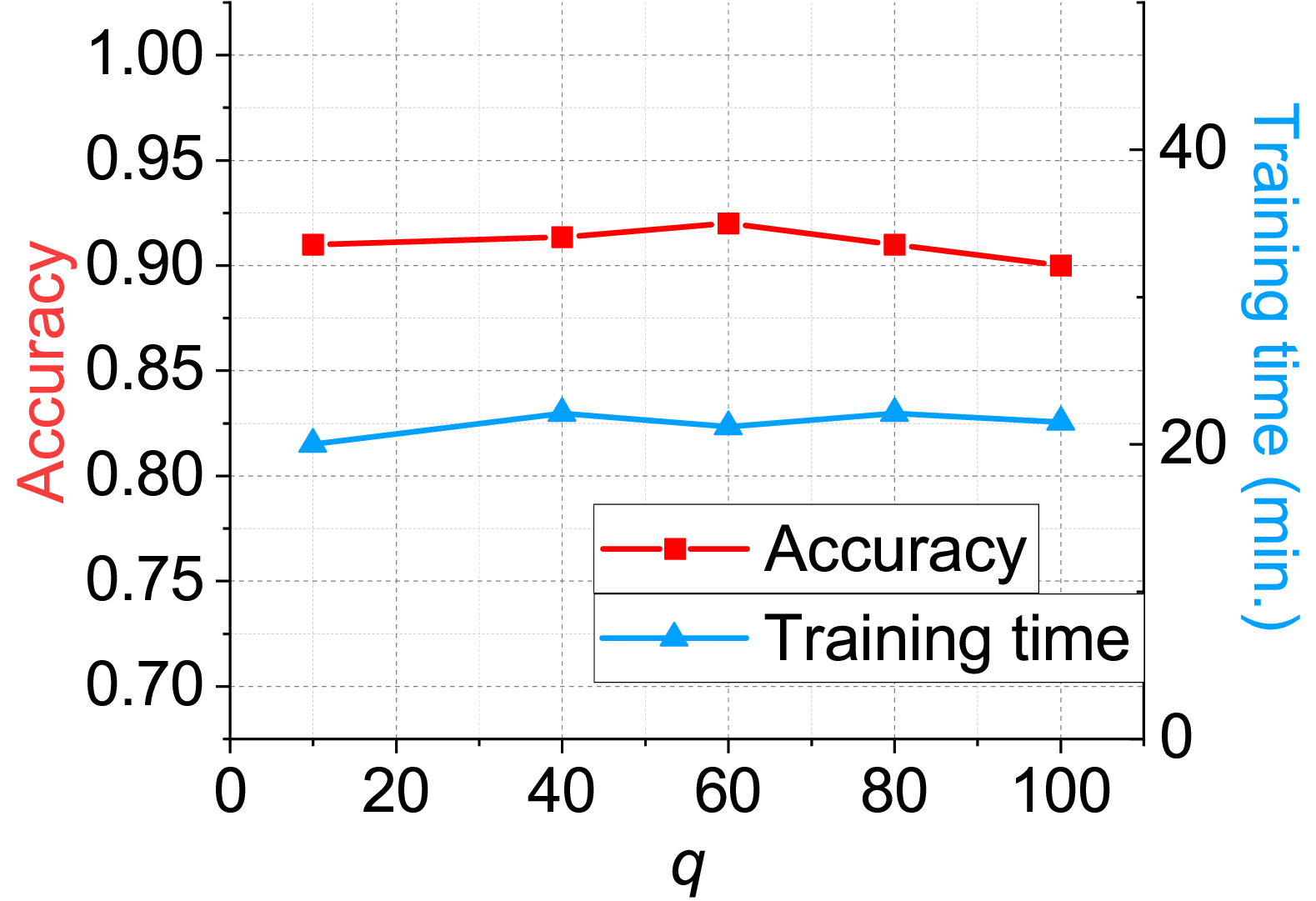,height=0.9in,width=1.05in}}
	\caption{The results of parameter analysis.}
	\label{fig:Parameters}
\end{figure}

We present a study of parameters, i.e., the number of classes in the training dataset, the size of training data set, and the value of $q$. When we test one parameter, the other parameters are fixed. Note that we just show the ResNet18 results on the CIFAR10 dataset. Similar results are also observed for other models or datasets. 

Figure \ref{fig:Parameters} (a) describes accuracy for different sizes of classes, and Figure \ref{fig:Parameters} (b) describes accuracy for different sizes of data in the training dataset. 
The results show that these factors have a relatively small impact on \textit{EncryIP}. 
Figure \ref{fig:Parameters} (c) shows accuracy for different values of $q$. As mentioned before, $q$ is a parameter related to security of the encryption scheme. In fact, the bigger $q$ is, the more security \textit{EncryIP} has and the lower the efficiency is. The results show that \textit{EncryIP} can maintain a stable trend on different values of $q$.




\end{document}